\documentclass[twocolumn]{aastex631}
\usepackage{multirow}
\usepackage{graphicx}
\usepackage{amsmath}

\graphicspath{{./}{figures/}}

\shorttitle{Swift J1728}
\shortauthors{Balakrishnan et al.}

\submitjournal{ApJ}

\begin{document}

\title{The Black Hole Candidate Swift J1728.9$-$3613 and the Supernova Remnant G351.9$-$0.9}

\correspondingauthor{M.~Balakrishnan}
\email{bmayura@umich.edu}

\author[0000-0001-9641-6550]{Mayura Balakrishnan}
\affil{Department of Astronomy, The University of Michigan, 1085 S. University Ave., Ann Arbor, MI, 48109, USA}

\author[0000-0002-2218-2306]{Paul A. Draghis}
\affil{Department of Astronomy, The University of Michigan, 1085 S. University Ave., Ann Arbor, MI, 48109, USA}

\author{Jon M. Miller}
\affil{Department of Astronomy, The University of Michigan, 1085 S. University Ave., Ann Arbor, MI, 48109, USA}

\author{Joe Bright}
\affil{Department of Astronomy, University of California at Berkeley, 501 Campbell Hall \#3411, Berekeley, CA, 94720, USA}
\affil{Department of Physics, University of Oxford, Keble Rd, Oxford, UK}

\author{Robert ~Fender}
\affil{Department of Physics, University of Oxford, Keble Rd, Oxford, UK}

\author{Mason Ng}
\affil{MIT Kavli Institute for Astrophysics and Space Research,  70 Vassar St, Cambridge, MA, 02139, USA}

\author{Edward Cackett}
\affil{Department of Physics \& Astronomy, Wayne State University, 666 W Hancock St, Detroit, MI, 48201, USA}

\author{Andrew Fabian}
\affil{Institute of Astronomy, University of Cambridge, Madingley Rd, Cambridge, UK}

\author{Kip Kuntz}
\affil{NASA Goddard Space Flight Center, Greenbelt, MD, 20771, USA}
\affil{The Henry A. Rowland Department of Physics and Astronomy, Johns Hopkins University, 3701 San Martin Drive, Baltimore, MD, 20218, USA}

\author{James C.A. Miller-Jones}
\affil{International Centre for Radio Astronomy Research, Curtin University, GPO Box U1987, Perth, WA, 6845, Australia}

\author{Daniel Proga}
\affil{Department of Physics, University of Nevada at Las Vegas, NV, 89154, USA}

\author{Paul S. Ray}
\affil{U.S. Naval Research Laboratory, 4555 Overlook Ave., SW. Washington, 20375, DC, USA}

\author{John Raymond}
\affil{Center of Astrophysics at Harvard, 60 Garden St.,Cambridge, MA, 02138, USA}

\author{Mark~Reynolds}
\affil{Department of Astronomy, The University of Michigan, 1085 S. University Ave., Ann Arbor, MI, 48109, USA}
\affil{Department of Astronomy, The Ohio State University, 140 West 18th Avenue, Columbus, OH, 43210, USA}

\author{Abderahmen~Zoghbi}
\affil{Deparment of Astronomy, University of Maryland, College Park, MD, 20742, USA}
\affil{HEASARC, Code 6601, NASA/GSFC, Greenbelt, MD 20771}
\affil{CRESST II, NASA Goddard Space Flight Center, Greenbelt, MD, 20771, USA}




\begin{abstract}
A number of neutron stars have been observed within the remnants of the core-collapse supernova explosions that created them.  In contrast, black holes are not yet clearly associated with supernova remnants.  Indeed, some observations suggest that black holes are ``born in the dark'', i.e. without a supernova explosion.  Herein, we present a multi-wavelength analysis of the X-ray transient Swift~J1728.9$-$3613, based on observations made with Chandra, ESO-VISTA, MeerKAT, NICER, NuSTAR, Swift, and XMM-Newton.   Three independent diagnostics indicate that the system likely harbors a black hole primary.  Infrared imaging signals a massive companion star that is broadly consistent with an A or B spectral type.  Most importantly, the X-ray binary lies within the central region of the catalogued supernova remnant G351.9$-$0.9.  Our deep MeerKAT image at 1.28~GHz signals that the remnant is in the Sedov phase; this fact and the non-detection of the soft X-ray emission expected from such a remnant argue that it lies at a distance that could coincide with the black hole.  Utilizing a formal measurement of the distance to Swift~J1728.9$-$3613 ($d = 8.4\pm 0.8$~kpc), a lower limit on the distance to G351.9$-$0.9 ($d \geq 7.5$~kpc), and the number and distribution of black holes and supernova remnants within the Milky Way, extensive simulations suggest that the probability of a chance superposition is  $<1.7\%$ (99.7\% credible interval).  The discovery of a black hole within a supernova remnant would support numerical simulations that produce black holes and remnants, and thus provide clear observational evidence of distinct black hole formation channels.  We discuss the robustness of our analysis and some challenges to this interpretation.
\end{abstract}

\keywords{X-rays: stellar-mass black holes --- X-rays: binaries --- Supernova Remnants -- X-ray transients --- Radio observations}

%
\section{Introduction} \label{sec:intro}
While neutron stars have been detected within the remnants of the core-collapse explosion that created them \citep{shternin2011, heinz2013}, clear evidence of a black hole within the supernova remnant of a progenitor star has been elusive. It has been argued that Cygnus X-1, arguably the most famous black hole X-ray binary, formed without a supernova explosion \citep{mirabel2003}. The detection of a black hole in a supernova remnant would broaden our understanding of massive stellar evolution, by demonstrating an independent formation channel that also allows the creation of a remnant.  An understanding of black hole formation will help us better grasp the demographics of X-ray binaries and gravitational wave source populations, as well as the final stages of the lives of massive stars.

As many as three sources may be examples of black holes within supernova remnants.  SS~433 is an X-ray binary associated with the supernova remnant W50.  The combination of dust along the line of sight to SS~433 and a complex local environment give rise to very high obscuration.  This makes it extremely hard to constrain the mass of the compact object using the motions of its massive companion star.  Indeed, spectroscopy and timing of the compact object itself may be impossible.  Several estimates based on optical photometry and spectroscopy yield masses above the neutron star range \citep{blundell2008, kubota2010, cherepashchuk2018}.  An association between the black hole candidate MAXI J1535-571 and the remnant G323.7$-$1.0 has recently been claimed \citep{maxted2020}.  However, differences in the column of obscuring gas along the line of sight suggest that the black hole and its low mass stellar companion may lie behind the remnant \citep{saji2018, miller2018}.  If they are associated, a specialized evolutionary channel may be required to allow a low-mass companion to fill its Roche lobe without significant binary evolution \citep{maxted2020}.  Finally, the asymmetric supernova remnant W49B is relatively young, with an age of only $\tau \simeq 5000$~years, yet no cooling neutron star or pulsar has been detected within it.  This absence, a rare morphology, and unusual metal abundances within W49B leave open the possibility that the explosion formed a black hole \citep{lopez2013}. 

Core-collapse supernovae explosions leave behind a compact object, along with diffuse emission in the form of a supernova remnant. The shock produced by the explosion can heat material to temperatures that make it visible in X-rays. Gamma-rays can be produced by energetic electrons through inverse Compton scattering, or by energetic ions through collisions in dense gas that produce pions \citep{bykov2018}. Neutron stars and pulsars emit radiation on their own, while the detection of stellar mass black holes is contingent upon the black hole accreting material from a companion star and the source going into outburst. In this paper, we report that the transient X-ray binary Swift J1728.9$-$3613 likely harbors a black hole primary, consistent with results from \citet{saha2022}, and may be physically associated with the supernova remnant G351.9$-$0.9.  

On January 25th 2019, the Swift Burst Alert Telescope (BAT) triggered and located a new transient in outburst that was named Swift J1728.9-3613 \citep{barthelmy2019}.  Subsequent monitoring with MAXI suggested that the source is an accreting X-ray pulsar or a hard-state black hole X-ray binary, based on the time scale of the flux rise and the spectral hardness of the source \citep{negoro2019}.  Observations with NICER revealed a highly absorbed spectrum, in addition to band-limited variability below 1~Hz and a narrow quasi-periodic oscillation (QPO) at 5.5~Hz, consistent with type-B QPOs that are often seen in the soft-intermediate state of black holes \citep{enoto2019}.  The MeerKAT radio telescope also observed Swift J1728.9$-$3613, detecting the source six days after discovery \citep{bright2019}.  After a relatively short bright phase, this outburst of Swift J1728.9$-$3613 decayed and the source returned to quiescence by mid-2019.  No additional outbursts from the source have been reported in any wavelength.

An extensive analysis of the spectral and timing properties of Swift J1728 using Swift and NICER observations concludes that the system hosts a black hole \citep{saha2022}. The authors in \citet{saha2022} report further on the detected type-B QPOs and note a partial hysteresis in the RMS-intensity diagram that supports the conclusion that Swift J1728 is a black hole X-ray binary.

The position of Swift~J1728.9$-$3613 \citep{barthelmy2019} is spatially coincident with  supernova remnant G351.9$-$0.9, previously discovered in radio observations \citep{mostcat}.  G351.9$-$0.9 is largely unremarkable: it lacks the spiral cocoon structure of W50 or the barrel structure of W49B.  Rather, G351.9$-$0.9 has a defined edge with heightened emission, and fainter, smoothly varying emission throughout its interior.  Like other remnants that are highly obscured by foreground dust and gas, it is not detected in optical wavelengths. The remnant is also not detected in the Fermi-Lat SNR catalog \citep{acero2016}. 


In Section 2 of this paper, we describe our observations and initial data reduction.  In Section 3, we analyze the nature of the X-ray binary Swift J1728.9$-$3613.  Section 4 describes the nature of the spatially coincident supernova remnant, G351.9$-$0.9. In Section 5, we investigate the probability of coincidental overlap between Swift J1728.9$-$3613 and G351.9$-$0.9. In Section 6, we discuss the strengths and weaknesses of our analysis, and the implications of an association between Swift J1728.9$-$3613 and G351.9$-$0.9.

\section{Observations and Preliminary Data Reduction}
We made observations of Swift~J1728.9 in outburst using Chandra, MeerKAT, NICER, NuSTAR, and Swift.  After Swift~J1728.9 entered quiescence, we also made X-ray observations of G351.9$-$0.9 using Swift and XMM-Newton, ensuring that the flux from a bright point source did not obscure or distort any diffuse emission from the remnant.  Radio observations with MeerKAT were made while Swift~J1728.9$-$3613 was in outburst, but this did not impact imaging of G351.9$-$0.9. NICER observations were used to construct a color-color diagram, and all NICER observations were searched for specific phenomena; for brevity, these are not listed within the table.
The key observations in our campaign are listed in Table \ref{tab:obsids}. 

\subsection{Chandra} \label{sec:chandra}
We observed Swift J1728.9$-$3613 with Chandra on six occasions.  In this analysis, we have focused on just two of these; a full spectroscopic analysis of the remaining observations will be the subject of a subsequent paper.  The first five Chandra observations were made using the High Energy Transmission Grating Spectrometer (HETGS), dispersed onto the ACIS-S array of CCDs.  The highest flux was observed in the first observation, OBSID 21213, obtained on 2019 February 02 (MJD 58526.4).  We focused on this spectrum to determine the column density and distance to Swift~J1728.9 (see Section \ref{sec:bh_dist}).  The final Chandra observation of Swift J1728.9$-$3613 was obtained when the source was faint, allowing us to use a standard imaging mode.  In OBSID 22289 (obtained on 2019 July 28, or MJD 58692.21), the HETG was not deployed, but the ACIS-S3 chip was operated in a 1/8 sub-array to mitigate photon pile-up \citep[e.g.][]{davis2001}, reducing the nominal frame time from 3.2~s to 0.4~s.  Both of these observations were reduced using the standard tools and procedures available in the CIAO suite, version 4.13, and the associated calibration files.  Prior to analysis, all observations were reprocessed using the CIAO tool \texttt{chandrarepro}.  Further details regarding our preparation and analysis of these observations follow below.

\subsection{NuSTAR}\label{sec:nustar}

Swift J1728.9-3613 was observed with NuSTAR on 2019 February 03 (MJD 58517.62), near the peak of the outburst.  The net exposure of the observation was 16.7~ks.  A count rate of $\sim280~{\rm counts}~{\rm s}^{-1}$ was recorded in each of the two NuSTAR Focal Plane Module (FPM) detectors: FPMA and FPMB. These detectors deliver both imaging and spectroscopy; however, owing to the brightness of the source, nearly the entire detector plane is illuminated and any weak diffuse emission is overwhelmed.  Using the standard routines in HEASOFT version 6.28 through the NuSTARDAS pipeline v2.0.0 and CALDB version 20210524, we extracted time-averaged spectra from the entire 3--79~keV pass band from the two NuSTAR detectors, using circular extraction regions of radius 120'' centered on the position of the source.  We extracted the background rate from regions of the same size as the source regions, placed at the edges of the detectors.  The full details of the observation and our analysis of the spectra is the subject of companion work by Draghis et al.\ (2023).  In the sections that follow, we utilize the key results of that analysis.

\subsection{XMM-Newton}

After Swift~J1728.9$-$3613 entered quiescence, XMM-Newton observed G351.9$-$0.9 on three occasions (see Table \ref{tab:obsids}). The first observation, OBSID 0860140101, was an 18~ks snapshot that started on 2020 September 04 (MJD 59116.17).  The next two observations were much deeper, each lasting 83~ks; OBSID 0901540101 started on 2022 February 28 (MJD 59638.07), and was immediately followed by OBSID 0910540201 starting on 2022 March 04 (MJD 59642.06).  

The central CCDs in the EPIC-MOS cameras are able to cover most of G351.9$-$0.9, potentially enabling an image with minimal disruptions owing to chip gaps and other considerations.  In all three observations, then, the PN and MOS cameras were run in ``full frame'' mode, providing the largest possible field of view.  The ``thin'' EPIC optical blocking filter was utilized in the initial snapshot, whereas the ''medium'' filter was used in the deeper exposures.  Owing to the large column density along this line of sight, our analysis was restricted to the EPIC cameras.  All of the data were reduced using XMMSAS version 19.1.0 and the corresponding ``CCF'' calibration files.  The specifics of how the data were prepared and analyzed for different purposes are detailed in the sections that follow.

\subsection{NICER}\label{sec:nicer}

The NICER telescope's primary instrument, the X-ray Timing Instrument (XTI), is made up of 52 live detectors with an operating range of 0.2--12 keV.  NICER observed Swift J1728.9$-$3613 at a high cadence throughout its 2019 outburst. For brevity, we list in Table \ref{tab:obsids} only the 48 observations featured in Figure \ref{fig:colorcolor}, although a spectroscopic analysis was performed on the initial 80 observations.  In order to conduct a search for Type-1 X-ray bursts in the light curve, and to search for coherent pulsations, we constructed and analyzed light curves from {\it all} of the NICER observations of Swift J1728.9$-$3613 available in the public archive.  All of the observations were processed using the standard pipeline tool, $\texttt{nicerl2}$ in HEASOFT version 6.28, and the files in CALDB version 20210728.  We utilized three background estimation tools that are publicly available.  Additional details follow in the sections below.

\subsection{Neil Gehrels Swift Observatory}
\label{sec:swift}
Swift observed Swift J1728.9-3613 six times: once in windowed timing (WT) mode and five times in photon counting (PC) mode (see Table \ref{tab:obsids}).  We processed all of the observations using tools available in HEASOFT version 6.28 and the associated calibration files.  Initial event cleaning was carried out using the task \texttt{XRTPIPELINE}.  Using the standard tools, we constructed light curves for the purpose of searching for X-ray bursts. In addition, Swift observed G351.9$-$0.9 on 2022 September 2022 (MJD 59844.56) for 1.5 ks.  This observation was obtained in PC mode and was also processed using the standard tools.

\subsection{MeerKAT}

We observed the field of Swift J1728.9$-$3613 with the MeerKAT radio telescope \citep{mauch2020} as part of the ThunderKAT large survey project weekly XRB monitoring program \citep{fender2016} beginning on 2019 February 12 (MJD 58533). We obtained a total of 13 observations amounting to an on-source time of 255 minutes. We initially reduced and imaged each observation independently using the \textsc{oxkat} \citep{oxkat2020} reduction pipeline, a set of semi-automated scripts that perform phase reference calibration, amplitude and phase self-calibration, and wide-field imaging \citep{mcmullin2007,offringa2014,kurtzer2017,kenyon2018,makhathini2018}. From each of these images we measured the flux density from G351.9$-$0.9.

In addition to imaging each observation individually, we also created a deep image of the field using the entire 255 minutes of on-source observing time. We concatenated the individual self-calibrated measurement sets in CASA and imaged the combined data using \textsc{WSClean} \citep{offringa2014}. We cleaned the field using $8\sigma$ automasking, then re-imaged the field using a custom mask based on the initial clean image. We employed multiscale cleaning to attempt to account for extended structure in the field and set a stopping threshold of 20$\mu\rm{Jy}$. This is significantly larger than the expected thermal noise from a 255 minute MeerKAT observation, however extended structures in the field make reaching the thermal noise impossible. 
\section{X-ray Binary Swift J1728.9--3613}
\subsection{Source Position}
After its discovery with the Swift/BAT \citep{barthelmy2019}, follow-up observations with the XRT recorded a source position of RA,DEC $=$ 17h 28m 58.64s, -36$^{\circ}$ 14' 37.7'' (J2000), with an error of 1.7'' \citep{kennea2019}.  This places the source within the region of the sky occupied by the catalogued supernova remnant G351.9$-$0.9 \citep{mostcat}.  In order to determine the best Chandra position of the source, 
we ran the CIAO tool \texttt{wavdetect} on Chandra OBSID 22289.  Swift J1728.9$-$3613 is detected at (RA,DEC) = 17h 28m 58.76s, -36$^{\circ}$ 14' 36.20''.  The statistical error on this position is negligible, but a systematic error of 0.3'' is reasonable based on the limits of Chandra's attitude control system \citep[e.g.][]{miller2017}.  The nominal positions differ by 2'' but are formally consistent within errors.  We adopt the Chandra measurement as the position of Swift~J1728.9 in the work that follows.
\subsection{Column Density and Distance}\label{sec:bh_dist}
The high column density along the line of sight to Swift~J1728.9$-$3613 makes optical measures of its distance impossible.  If the source is once again observed in radio during a prolonged outburst, it may be possible to determine its proper motion and distance using very long baseline interferometry.  In the meantime, the same column density that complicates other distance measures can be turned into a distance constraint in X-rays.  In short, the distance to a given source can be determined by integrating the gas density distribution along the line of sight until the accumulated column density matches that measured in X-rays.  Using a combination of measured galactic $n_H$ radial density profiles \citep{kalberla2009, marasco2017}, we obtained a continuous distribution of neutral hydrogen in the galactic midplane as a function of galactocentric radius.  For a given line of sight, we calculate the galactocentric radius and get the equivalent hydrogen number density, $n_H$ (cm$^{-3}$), and integrate along that line of sight to obtain a column density $N_H$ (cm$^{-2}$).  

To determine the column density to Swift~J1728.9$-$3613, we made joint fits to the combined MEG and combined HEG spectra obtained in Chandra OBSID 21213 using XSPEC version 12.11.1 \citep{xspec}.  Prior to fitting, the spectra were binned according to the ``optimal binning'' algorithm detailed in \citet{kaastra2016}, using the tool \texttt{ftgrouppha}.  Fits to the MEG spectrum were restricted to the 1--5~keV range, and fits to the HEG spectrum were restricted to the 1.2--8~keV range, owing to poor sensitivity below and above these bounds.  The spectra were fit minimizing the Cash statistic \citep{cash1979}, allowing a normalizing constant to float between them.  The \texttt{tbabs} \citep{wilms2000} model was used to describe line-of-sight absorption, after adopting the required cross sections and abundances (``Verner'' abundances and ``Wilms'' cross-sections, within XSPEC).  The continuum is well described with only a disk blackbody \citep{mitsuda1984}.  This model measures $N_{H} = 4.22\pm 0.02\times 10^{22}~{\rm cm}^{-2}$, $kT = 1.27\pm 0.01$~keV, and a flux normalization of $K=242\pm 5$, and gives a fit statistic of $C = 3270.4$ for $\nu = 2365$ degrees of freedom ($1\sigma$ errors).  The constant indicates that the MEG normalization is 8\% higher than the HEG in OBSID 21213.

This value of the column density to Swift~J1728.9$-$3613 is broadly consistent with that measured using NuSTAR, $N_{H} = 3.7^{+0.2}_{-0.1} \times 10^{22} ~{\rm cm}^{-2}$ (Draghis et al.\ 2023).  In that work, they considered a series of more sophisticated models, including thermal emission from the accretion disk, non-thermal emission from the corona, and reflection from the innermost region of the disk.  These measurements are not in formal agreement at the $1\sigma$ level, but the degree of agreement is excellent considering the different pass bands and point spread functions of each telescope, and the different models required to fit the spectra.  

Integrating along the line of sight, the column density measured with Chandra nominally translates to a distance of $d = 9.08^{+0.11}_{-0.09}$~kpc, and the column density measured with NuSTAR indicates a distance of $d = 7.9^{+0.34}_{-0.27}$~kpc.  Although it is likely the case that Chandra is better suited to this measurement, we conservatively regard the difference as a systematic error, and adopt a distance constraint of 7.6~kpc $\leq d \leq$ 9.2~kpc, or $d = 8.4\pm 0.8$~kpc.  

In order to verify that this technique delivers reasonable distance estimates, we collected published column densities from a number of X-ray binaries located close to the plane of the Milky Way that have independent distance constraints.  For each source, we calculated the distance to the source based on its column density, and compared that value to the independent constraint.  In general, the two are in good agreement; at least within the plane, it appears that our technique is robust.  Please see Appendix A for additional details.

\subsection{X-ray Colors}
\label{sec:xray colors}

\begin{figure*}[!t]
    \centering
    \includegraphics[width=0.95\textwidth]{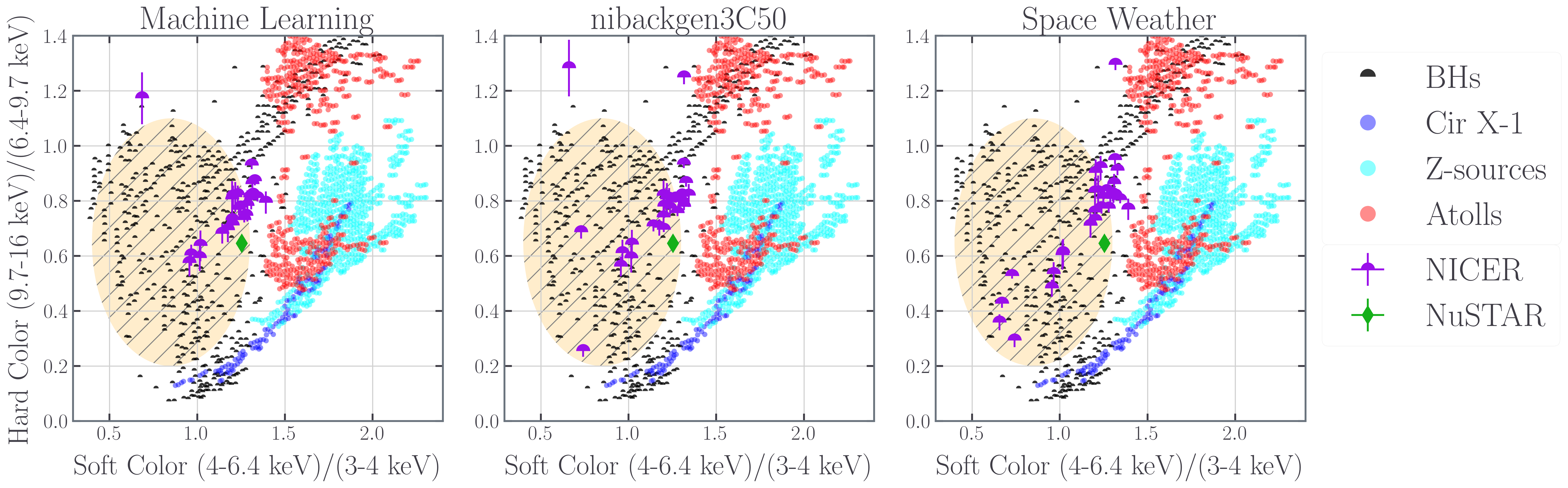}
    \caption{Recreations of Figure 8 from \citet{done2003}, where black holes are represented by black half-circles, and the filled circles refer to neutron stars. The red, cyan, and blue filled circles correspond to atolls, Z sources, and Circinus X-1, respectively. The orange hatched region represents the area on the plot inaccessible by neutron stars due to the physical boundary at their surfaces. We overplot values from our analysis: the violet half-circles are measured from NICER spectra, while the green diamond point represents the NuSTAR observation. Each panel corresponds to NICER hard and soft color values derived using different background estimation routines. Please see the text for details.}
    \label{fig:colorcolor}
\end{figure*}
Although accreting neutron stars and black holes both have accretion disks and non-thermal coronae, the build-up of accreted material in a boundary layer on the neutron star surface is potentially a distinguishing feature. \citet{done2003} analyzed a large set of RXTE observations of X-ray binaries to show that the {\em absence} of boundary layer emission may be used to infer that the compact object in a given X-ray binary is a black hole, and that these differences are evident in X-ray colors.  In that work, no neutron star systems show a hard color (defined as the ratio of 9.7 - 16 keV flux to 6.4 - 9.7 keV flux) between 0.25 and 1, and a soft color (defined as the ratio of 4 - 6.4 keV flux to 3 - 4 keV flux) that is lower than $\sim$ 1.4.   For reference, we have recreated most of Figure 8 from \citet{done2003} in Figure \ref{fig:colorcolor}, with black semi-circles indicating black hole systems, filled circles indicating neutron star systems, and the orange hatched region signifying X-ray color ranges inaccessible to neutron stars. We do not replicate the tracks shown in the original figure that arise from models with different combinations of disk temperature and photon index, as these are not relevant to this work.
NICER obtained the densest monitoring of Swift J1728.9$-$3613, and its effective area is much higher than that of Swift close to 10~keV.  In order to apply the diagnostics developed by \citet{done2003}, we therefore analyzed the first 50 NICER observations of Swift J1728.9$-$3613.  These represent the brightest observations, from which the most robust colors can be measured.  For each observation, we filtered for events in 1--10 keV energy range and removed instrumental flaring intervals.  Source spectra were extracted from cleaned event files, and the background spectra were generated using $\texttt{nibackgen3C50}$ \citep{remillard2021}.  The standard NICER response matrix and ancillary files were used from CALDB version 20210728.  Prior to fitting each spectrum in XSPEC, the data were binned using the ``optimal binning'' scheme developed by \citet{kaastra2016} and implemented in the tool \texttt{ftgrouppha}.  Each spectrum was fit over the 1--10~keV range with a simple model consisting of a disk blackbody and a power-law, modified by line of sight absorption: \texttt{tbabs $\times$ (diskbb + powerlaw)}.  
The hardest band in the \citet{done2003} diagnostic, 9.7--16.0~keV, extends above the NICER pass band.  Since this scheme relies on fluxes, not count rates, it is possible to extrapolate models into this regime.  However, caution is required, because the NICER background is complicated by variable space weather, its position aboard the International Space Station, and the fact that NICER is not a focusing instrument.  Background flux is inevitably highest at the edges of an instrumental pass band.  We therefore repeated the above procedure, using two alternative derivations of the NICER background, the ``space weather'' method (also supported by the NICER team) and \texttt{nicer-background} v0.4.t1.200e (Zoghbi et al. in prep; \url{https://github.com/zoghbi-a/nicer-background}). We grouped the spectra using oggrpha (\url{https://github.com/zoghbi-a/aztools/blob/master/scripts/ogrppha.py}), which allows us to take into account the high background by binning the spectra to a minimum S/N of 3 and oversampling the detector resolution by a factor of 3.
Again following the work of \citet{done2003}, we used the counts in each band as a proxy for the errors, allowing the fact that we were using 3 - 10 keV spectra to be extrapolated up to 16 keV to be conveyed in the errors on the hard color. To illustrate, the error on the soft color was calculated as follows:
\[ \frac{\sigma_{soft}}{\text{color}_{\rm soft}} = \sqrt{ \left(\frac{\sqrt{ \text{counts}_{\rm 3 - 4 keV}}}{\text{counts}_{\rm 3 - 4 keV}} \right)^2 + \left(\frac{\sqrt{ \text{counts}_{\rm 4 - 6.4 keV}}}{\text{counts}_{\rm 4 - 6.4 keV}} \right)^2   } \] 
Figure \ref{fig:colorcolor} shows the location of these NICER observations of Swift J1728.9$-$3613 on the color--color diagram with violet half-circles. The NuSTAR observation is indicated by the green diamond. The three panels in the figure correspond to the three different background methods that were utilized.  For clarity, only those observations yielding colors with fractional errors less than 10\% are plotted (29 of 50 observations).  The strong majority of the points fall within the zone that \citet{done2003} identify as only possible for black holes.   Including the observations with less certain colors does not change this result.  Similarly, the results are robust against the implementation of different binning schemes within the simple spectral fits.  

The observation of Swift J1728.9$-$3613 made with NuSTAR is particularly important to our analysis, and the NuSTAR pass band is more similar to the RXTE spectra considered by \citet{done2003}.  Adopting the simple \texttt{tbabs*(diskbb + powerlaw)} model, fitting the full 3--79~keV pass band, and deriving errors based on the counts in each band, we derive a hard color ratio of 0.652 $\pm$ 0.004 and soft color ratio of 1.258 $\pm$ 0.006.  This point is shown in Figure \ref{fig:colorcolor} as a green diamond that also lies within the zone that is only accessible to black hole X-ray binaries.  Draghis et al.\ (2023) fit the NuSTAR spectrum of Swift~J1728.9 with a set or disk reflection models; the hard and soft colors derived from these models differ from the above values by less than 1\% and still fall within the black hole zone.

\subsection{Spin Parameter}

The dimensionless angular momentum or ``spin parameter'' of a compact object, $a = cJ/GM^{2}$, can act as a direct probe of its nature.  Neutron stars have an observed spin limit of approximately $a \simeq 0.2$ \citep{millermiller2015, reynolds2021}, while the theoretical upper limit on neutron star spin due to break up is $a \leq 0.7$ \citep{2011ApJ...728...12L}.  Previous timing and spectral analysis \citep{saha2022} and our measurements of X-ray colors observed in Swift J1728.9$-$3613 are suggestive of a black hole primary, but a spin parameter that demands a black hole primary is potentially more definitive.

X-ray measures of black hole spin all rely on the premise that the accretion disk obeys the innermost stable circular orbit (ISCO) set by the spin parameter, when the Eddington fraction is moderate.  The most broadly applicable method of determining black hole spin is based on measuring relativistic distortions to the reflection spectrum of the inner accretion disk.  The extremity of these distortions is set by the ISCO, and therefore by the spin of the black hole.  This is now a well-established spectral modeling methodology that has resulted in numerous precise spin measurements, especially when the broad pass band and high energy sensitivity of NuSTAR is utilized \citep[see, e.g.][]{draghis2022}.

In a companion work, Draghis et al.\ (2023) modeled the relativistic reflection features present in the NuSTAR observation (OBSID 90501303002).  Fits to the 3--79~keV NuSTAR spectra with the ``relxill'' family of disk reflection models \citep[e.g.][]{dauser2013, garcia2013} yield a black hole spin parameter of $a = 0.86\pm 0.02$ ($1\sigma$ statistical errors).  This value nominally excludes the theoretical maximum for neutron stars at the $5\sigma$ level of confidence.  The highest observed neutron star spin parameters are excluded at much higher levels of confidence.  

This spin measurement is a lower limit; determined by the innermost radius consistent with the observed spectrum. Inner regions of the disk that are not illuminated would not be measured, and therefore the spin of the black hole could be higher.
\subsection{Search for Pulsations}
Although Swift~J1728.9$-$3613 appears to harbor a black hole, the detection of coherent pulsations would unambiguously signal a neutron star primary.  For this reason, it is important to understand the sensitivity of the NICER data to coherent pulsations similar to those observed in other X-ray binaries.  These dense monitoring obtained with NICER, and its pass band, afford the best chance of detecting pulsations.

\begin{figure}[!t]
    \centering
    \includegraphics[width=0.38\textwidth, angle=270]{ 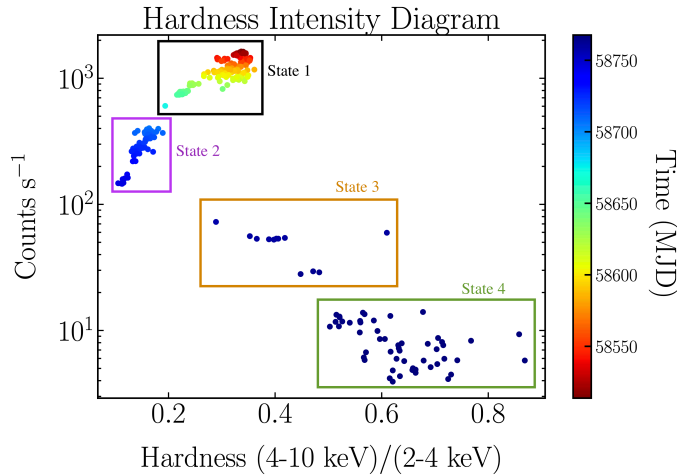}
    \caption{Hardness intensity diagram from NICER observations of Swift J1728.9$-$3613. The different boxes show how the observations were split into four spectral states in order to do the pulsation search.  No pulsations are found, with upper limits ruling out pulse amplitudes in the range typically detected in neutron star X-ray binaries.}
    \label{fig:hid}
\end{figure}

Figure \ref{fig:hid} shows a hardness--intensity diagram, constructed using NICER observations obtained in between MJD 58513--58548, 58559--58613, 58633--58650, and 58651--58768.  Specifically, the total count rate is plotted versus the ratio of counts in the 4--10~keV to 2--4~keV band.  In this HID, the data can be grouped into four spectral states.  The time series was broken up into 128 second segments, and for each segment, we constructed 1 s-binned light curves; if the light curve had a duty cycle of at least 90\%, it was admitted into the calculation of the averaged spectrum.   

For each admitted 128~s segment of data from these observations, the time series was then binned to 0.5 ms (i.e., Nyquist frequency of 1000 Hz), and we calculated the Leahy-normalized power spectrum.  The total power spectrum is then averaged across all of the admitted segments \citep{bartlett48}.   In the end, we had 152, 17, 182, and 195 segments corresponding to the four spectral states identified for the search.  No coherent pulsations were detected above the noise.  

We then calculated the 3-sigma upper limits on the fractional pulsation amplitude using the formalism laid out in \citet{vaughan1994}, modifying it for the case of a stacked power spectrum.  The maximum possible pulse amplitudes (3-$\sigma$ upper limits) for Swift J1728.9$-$3613 at 10, 100, and 400 Hz are 0.0051\%, 0.0078\%, 0.36\%, and 0.053\%, respectively. This is lower than typical pulse amplitudes for millisecond pulsars, which range from 1-10 \% \citep{patruno2021}.

Acceleration searches are excellent for detecting pulsations from pulsars in binary systems, because the Doppler motion due to the orbit modulates the NS pulsation frequency, and the signal that was spread across multiple Fourier bins can be combined \citep{presto}.
We initially `bunched up’ adjacent GTIs; that is, if the interval between the end of one GTI and the start of the next GTI is sufficiently short, the two GTIs were combined. With a gap of 10 s, we had 301 GTIs in the end, and we only performed acceleration searches on GTIs that were longer than 200 s (175 such GTIs). The acceleration searches were done with \texttt{accelsearch} in PRESTO over 0.01--100 Hz \citep{presto}. We also set $z_{\rm max}$ to 100, where we surmised that any Doppler-modulated signal would have been smeared across a maximum of 100 Fourier bins (standard choice in the field). The range of accelerations searched is given by $a = zc/(f_0 T^2)$ where $a$ is the orbital acceleration (${\rm m\,s^{-2}}$), $z$ is the number of Fourier bins the signal would smear across, $c$ is the speed of light (m/s), $f_0$ is the pulsar frequency (Hz), and $T$ is the span of the segment being searched. Upon visual inspection of the candidates and their diagnostic plots (e.g., subintegration plots), there were no significant candidates that could be identified as a coherent signal.

\subsection{A Search for Type I X-ray Bursts} \label{sec:bursts}


Type I bursts are thermonuclear runaway events that result from the build-up of accreted material on the surface of a neutron star.  Therefore, the detection of one or more bursts in a given X-ray binary unambiguously signals that it harbors a neutron star.  No bursts have been reported in Swift~J1728.9$-$3613, but it is important to make a systematic search and to understand whether or not prior observations were capable of detecting bursts with the properties and recurrence times observed in other sources. It is also important to note that not all neutron star binaries display type-I bursts.
 
The frequency, strength, and timescales of Type I bursts depend on the composition of the burning material and its metallicity. There are variations from source to source because of differences in the core temperature of neutron stars and the accretion rate \citep{galloway2008}.  In particular, Type I bursts are typically seen at Eddington fractions of $\lambda_{\rm Edd} = 0.003-0.3$ when looking at the 2.5 -- 25 keV flux band, \citep{galloway2008}.  

Using the distance estimate to Swift J1728.9$-$3613 determined in this paper ($d = 8.4 \pm 0.8 \rm \ kpc$) and the spectra extracted from all of the observations made using Chandra, Swift, NICER, and NuSTAR, we selected observations that fall within this window (extrapolating the pass band, where necessary). $L_{edd}$ was calculated with an assumed mass of 1.4 M$_\odot$. NICER observations dominate this set, owing to the dense monitoring it achieved across the outburst of Swift~J1728.9$-$3613.  As noted above, NICER has a complex background; we therefore generated background light curves in the 0.4–15 keV band from telescope environment data and compared them to the source light curves.  This was repeated for all three background formulations.  All potential bursts were due to increases in the detector background, and not due to flaring or bursting in Swift~J1728.9$-$3613.  We note that we repeated this procedure for all NICER observations of Swift J1728.9$-$3613, without respect to Eddington fraction, and we again find no evidence of bursts from the source.  In their timing analysis, \citet{saha2022} also report no bursts.

\subsection{The Binary Companion Star}

The Chandra and NuSTAR spectra of Swift~J1728.9$-$ 3613 in outburst are broadly consistent with a column density of $N_H = 4 \times \rm 10^{22} \ cm^{-2}$.  The color excess is given by: $E(B-V) = \frac{N_H}{5 .8 \times 10^{21} \rm atoms \ cm^{-2} mag^{-1}} $ \citep{allens}.  For typical conditions in the ISM of the Milky Way, the optical extinction, A$_V$, is related to the color excess by: $\frac{A_V}{E(B-V)} \sim 3.1$.  Our column density measurements then correspond to A$_V \sim 21$ magnitudes of extinction.  Even an A0 supergiant with M$_V$ = $-$6.5 would then have an apparent magnitude $\sim$ 29-30.  This is at the very limit of what can be detected in the deepest observations with Hubble; it is well beyond what can be detected with the Swift UVOT telescope and ground-based telescopes.  We conclude that the optical counterpart of Swift~J1728.9$-$3613 has not been detected, and that optical observations cannot reveal the nature of the companion star and binary system.

\subsection{Infrared Observations}
\begin{figure}[!t]
    \centering
    \includegraphics[width=0.45\textwidth]{ 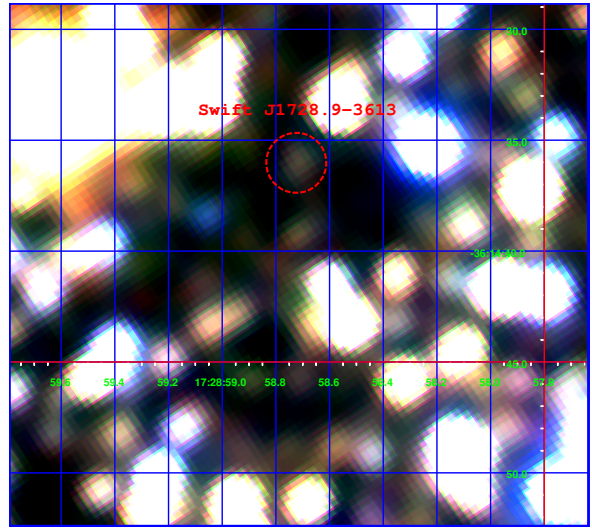}
    \caption{Finder chart showing the IR emission from ESO-VISTA of Swift J1728. The red circle represents a region of radius equal to 1.3 arc seconds, centered at the known position of the source. The image was composed by displaying the $K_s$ band in red, the $H$ band in green, and the $J$ band in blue.}
    \label{fig:finder_chart}
\end{figure}

\begin{figure*}[!t]
    \centering
    \includegraphics[width=0.95\textwidth]{ 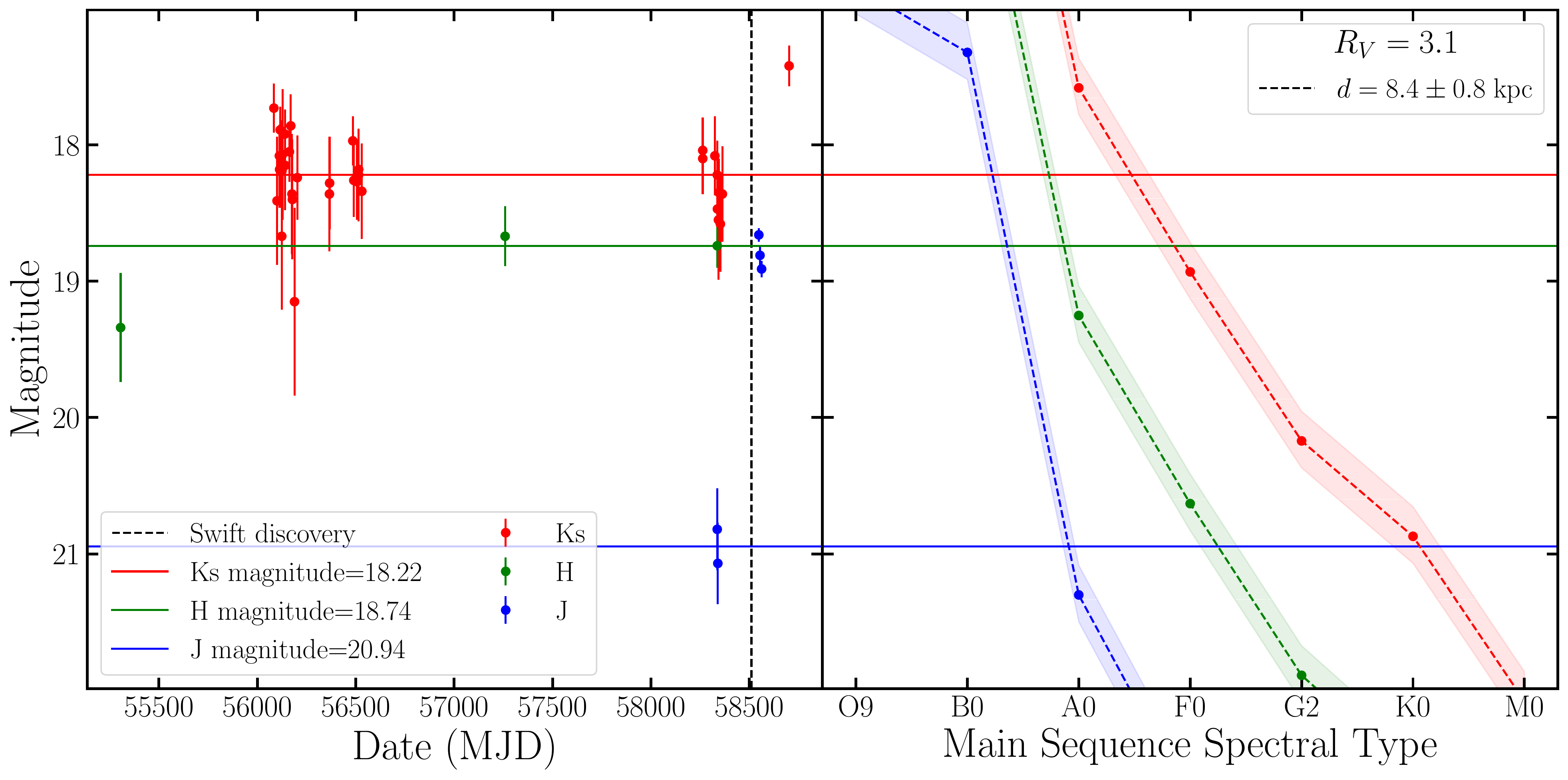}
    \caption{\emph{Left}: The IR light curve of Swift J1728. Red points represent $K_s$ magnitudes, green points represent $H$ magnitudes, and blue points represent $J$ magnitudes. The horizontal lines show the median magnitude prior to the outburst, in the three bands. The vertical dashed line shows the date of the Swift detection of the outburst. 
    \emph{Right}: Apparent magnitudes for main sequence stars assumed to be at a distance of $8.4 \pm 0.8~\rm kpc$ (dashed line with shaded regions). Values are computed assuming $N_H=4\times10^{22}\;cm^{-2}$ and $R_V=A_V/E(B-V)=3.1$.}
    \label{fig:IR_analysis}
\end{figure*}

We examined archival infrared observations in $J, H,$ and $K_s$ bands from the VIRCAM camera on the ESO-VISTA telescope to search for emission at the Chandra X-ray position of Swift J1728.9$-$3613.  Figure \ref{fig:finder_chart} was composed using a $K_s$ frame to illustrate the color red, an $H$ frame to show color green, and a $J$ frame to show color blue. The red circle has a radius of 1.3'' and is centered at the Chandra position.  A candidate source is clearly distinguishable from the background.  
We therefore downloaded a total of 31 $K_s$ frames, 7 $J$ frames, and 3 $H$ frames (all fully reduced) from the ESO Science Archive and ran Gaussian-weighted aperture photometry at the position of Swift J1728.9$-$3613 and of roughly 4000 other point sources with known $J, H,$ and $K_s$ magnitudes obtained from the Two Micron All Sky Survey (2MASS) catalog \citep{skrutskie2006}.   We required the catalog sources to have an apparent J magnitude between $11.5\leq m_J \leq 14.5$, in order to avoid saturated sources and still have reliable magnitude uncertainty measurements. The aperture photometry was performed via the NOAO IRAF software suite, using circular extraction regions with 5-pixel radius. The Gaussian weighting was set to be equal to the FWHM of the individual images. We extracted the background from annuli centered at the position of the sources, with an inner radius and width of 10 pixels, and calculated the background rate as the mode of the values of the pixels in the annulus. In order to account for possible small calibration errors in the WCS registration of the frames, we allowed a centering algorithm to shift the position of the source and background regions by at most one pixel.  We excluded sources where a region shift of more than one pixel was needed.  Last, we performed differential photometry in order to calculate the apparent magnitude of the source at the position of Swift J1728.9$-$3613.
The left panel in Figure \ref{fig:IR_analysis} shows the apparent magnitude in the $K_s, H,$ and $J$ filters in red, green, and blue (respectively) as a function of time, over the past 11 years.  Non-detections and measurements with an uncertainty larger than 0.7 magnitudes were not plotted or used in subsequent analysis.  The dashed vertical black line corresponds to the time of the Swift discovery of the source.  The horizontal lines correspond to the median $K_s, H,$ and $J$ magnitudes over the measurements, prior to the 2019 outburst.  The median apparent magnitudes correspond to $m_{K_s}\sim 18.2$, $m_{H}\sim 18.7$, and $m_J\sim20.9$.  Shortly after the 2019 outburst, the $J$ apparent magnitude of the source is observed to be 2 magnitudes brighter than the median before the outburst; thereafter, the $J$ flux decays in time.  About 200 days after the Swift detection, the $K_s$ apparent magnitude is $\sim 0.7$ magnitudes brighter than prior to the outburst.  This suggests a significant IR brightening either coincidental to the outburst or shortly after that could be associated with jet production in harder spectral states or disk brightening, followed by a decay to the value observed 200 days after the Swift discovery. 
\begin{figure*}[!t]
    \centering
    \includegraphics[width=0.97\textwidth]{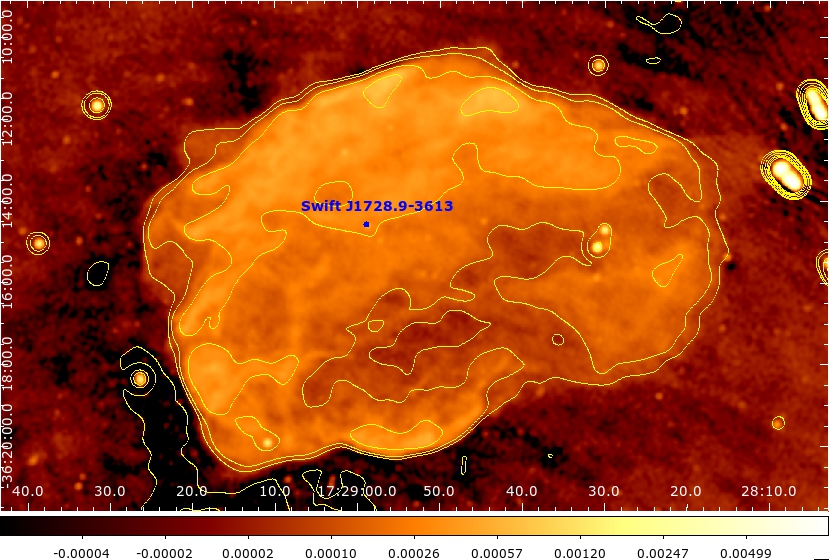}
    \caption{Stacked $1.28$ GHz MeerKAT image of G351.9$-$0.9, with white contours at 25 $\mu$Jy $\times$ (-3, 3, 6, 9, 12, 15) $\sigma$. The radio color bar indicates the flux density scale in Janskys. The Chandra position of Swift J1728.9-3613 is indicated with a filled blue circle.}
    \label{fig:snr_radio}
\end{figure*}
The right panel in Figure \ref{fig:IR_analysis} shows the expected $K_s, H$, and $J$ apparent magnitudes of main sequence stars of spectral types ranging from O9 to M0, in red, green, and blue respectively.  The dashed lines show expected apparent magnitudes for different stellar types (values taken from \citealt{allens}), calculated for a distance of $d=8.4 \pm 0.8$~kpc and a hydrogen column density of $N_H=4\times10^{22}$ cm$^{-2}$ while accounting for the gas-phase, the grain-phase, and molecular interstellar medium and a ratio of total to selective extinction $R_V=A_V/E(B-V)=3.1$. While the pre-outburst emission does not perfectly correspond to the expected magnitudes of a catalog main sequence star, the values of the apparent magnitudes in the three bands and the colors are broadly consistent with a late B or A class star. 
\section{Supernova Remnant G351.9--0.9}
\subsection{Radio Imaging}
The deep, 255-minute MeerKAT image of the field containing Swift~J1728.9$-$3613 yielded an excellent view of the supernova remnant G351.9$-$0.9 shown in Figure \ref{fig:snr_radio}.  The Chandra position of Swift~J1728.9$-$3613 clearly places the source within the central region of the remnant.  G351.9$-$0.9 has an elliptical shape, with a major axis that is approximately 12' long, and a minor axis that is approximately 9' long.  It is center-filled, but retains a clear edge structure with enhanced emission along its perimeter.  The appearance of G351.9$-$0.9 in the deep MeerKAT image resembles that of the Cygnus Loop, which is in the Sedov phase \citep{fesen2018}. Both remnants have higher brightness patches where the supernova blast wave is interacting with denser clouds and the shock has become radiative, increasing the radio emission due to the compression \citep{raymond2020}. This implies the remnant is still in the Sedov phase, and therefore less than $\tau  = 30,000$ years old \citep{cioffi1988}. A pressure-driven post-Sedov supernova remnant would appear more disjoined and more strongly limb-brightened, while a pre-Sedov remnant would be smaller and brighter given the distance constraint provided by the column density. At a distance of 8.4 kpc (coinciding with Swift J1728), the supernova remnant would have a diameter of $\sim$ 20-30 pc, which is reasonable for a remnant in the Sedov phase.


To constrain the total flux density contribution from the remnant, we used image analysis tools from CARTA \citep{carta}. In our final clean image we measure a sum of $64\,\rm{Jy~beam}^{-1}$ (corresponding to a flux density of $1.8\,\rm{Jy}$) from $\sim3.4\times10^{5}$ pixels from a manually drawn region encompassing the structure. Within the same region we measure a sum of $5.3\,\rm{Jy~beam}^{-1}$ (corresponding to a flux density of $0.2\,\rm{Jy}$) in the residual image, implying $\sim1.6\,\rm{Jy}$ within the sky model of the extended emission. 

\begin{figure*}[!t]
    \centering
    \includegraphics[width=0.97 \textwidth]{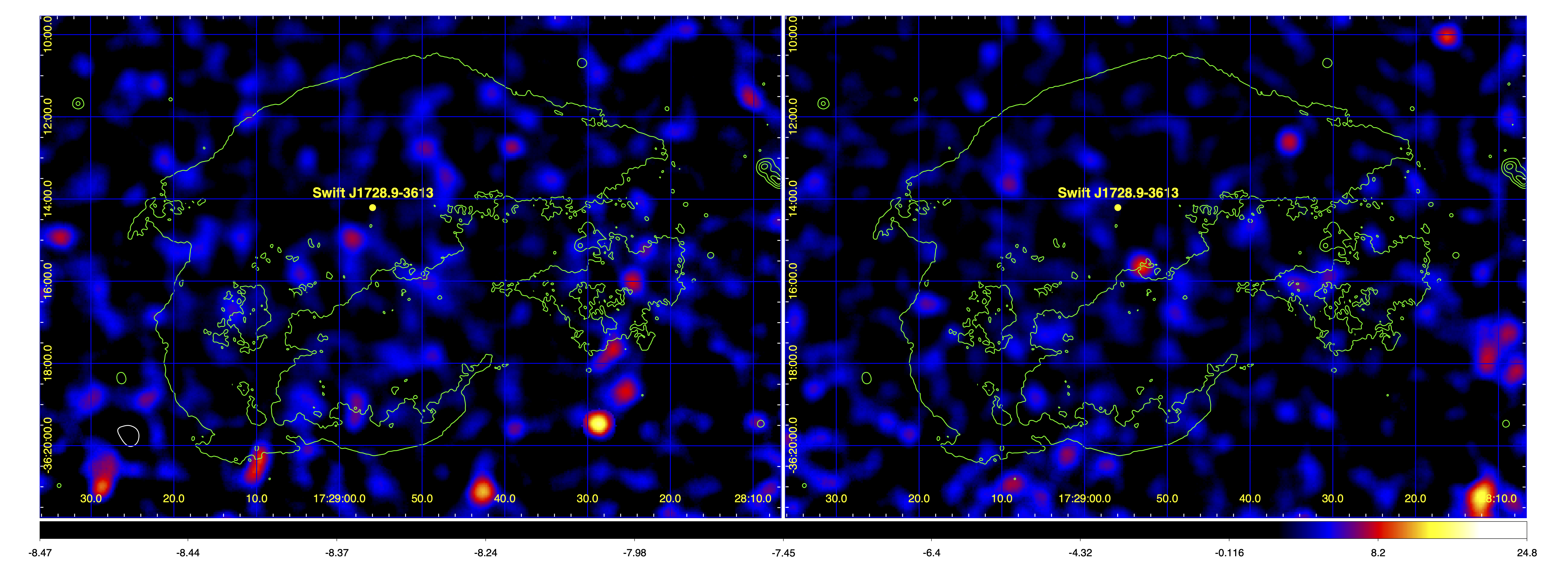}
    \caption{LEFT: The combined MOS-2 image of the field containing G351.9$-$0.9 in the 2.0-7.5 keV band.  RIGHT: The combined MOS-2 image of the same field in the 0.3-1.0 keV band.  In both cases, strong point sources were excised, soft proton flaring was removed on a chip-by-chip basis, an estimate of the quiescent particle background has been subtracted, and the images are exposure-corrected and then smoothed with a Gaussian ($\sigma =$10 pixels). The contours shown in green trace the outline of the remnant in radio bands, as detected with MeerKAT.  In both panels, the Chandra position of the black hole candidate Swift J1728.9-3613 is indicated with a filled yellow circle.  There is no evidence of X-ray emission from the remnant, potentially consistent with a spectrally soft, young remnant that is obscured by the relatively high foreground column density.}
    \label{fig:snr_xray}
\end{figure*}

There are a number of point sources embedded within the diffuse radio structure, including Swift J1728.9$-$3613. However, we did not remove them before calculating the total flux density as they each contribute $<<1\%$ of the total.  In addition to the diffuse structure, there are a number of bright extended structures detected outside of the remnant by MeerKAT; these cause the image noise to be highly non-Gaussian and non-uniform.  It is therefore difficult to determine if the remaining flux density in the residual image corresponds to the elevated noise level, or to genuine emission, or indeed if the clean model contains exclusively emission from the nebula itself.  We therefore place a tentative range of $1.6$-$1.8\,\rm{Jy}$ on the total diffuse emission. Our estimates are consistent with other reported measurements of G351.9-0.9: 10.7 Jy at 300 MHz \citep[uGMRT;][]{veena2019} and 2.0 Jy at 843 MHz \citep[MOST;][]{mostcat}. These measurements translate to $\sim$ 0.9 - 1.6 Jy at 1.28 GHz for the 1 $\sigma$ range of flux densities and spectral indices reported.

Additionally, we note that, due to MeerKAT being a radio interferometer, our images are affected by the `zero-spacing' problem. This results in flux density not being recovered beyond a certain angular scale determined by the shortest baseline length. Angular scales $\lesssim10\arcmin$ should be well recovered by Meerkat \citep{knowles2022} due to the short minimum baseline and abundance of short spacings. This is the approximate size of G351.9$-$0.9, so while our flux density measurement is formally a lower limit it should provide a reasonable estimate for the total emission from the remnant at $1.28\,\rm{GHz}$.

\subsection{X-ray Imaging}

The EPIC cameras have a very large field of view, approximately 30'.  This feature, and the sensitivity afforded by the collecting area of these cameras, make them excellent for surveys of point sources and bright extended sources.  However, imaging of faint extended sources is more difficult.  Instrumental flaring due to soft proton flux, the quiescent particle background of the instruments, and the diffuse astrophysical foreground caused by solar wind charge exchange all pose significant difficulties.  The ESAS software suite \citep{snowden2004, snowden2008} was developed to handle imaging of faint diffuse structures; it is particularly effective when a source is large enough that local backgrounds are difficult or impossible to obtain.  This suite combines standard SAS routines and new routines to produce images and spectra that are as free from these complications as possible.  We found that some of the tools within the ESAS suite are incompatible with recent SAS releases, and therefore utilized a development version of the ESAS suite that has been updated to work with the latest libraries and compilers.

G351.9$-$0.9 is slightly larger than the central CCD in each MOS camera, requiring the use of adjacent CCDs.  The loss of two of these CCDs in the MOS-1 camera prevents robust exposure correction and background estimation at the expected source boundary.  Similarly, while the PN camera remains the best instrument for point sources, the growing number of bad pixels and columns -- especially close to the chip boundaries that run through the center of G351.9$-$0.9 given its position within the field -- it is likely not suited to studies of faint extended emission.
In contrast, all of the CCDs in the MOS-2 camera continue to operate, and the detector is relatively free of bad pixels and columns.  Particularly since XMM-Newton does not dither, this is important to robust imaging of faint emission.  For these reasons, our analysis of the field containing G351.9$-$0.9 is restricted to the MOS-2 camera, although mosaics including the MOS-1 and PN cameras were constructed and consulted. The deeper mosaics using the more disparate data did not reveal any further insight.

We ran the standard SAS task \texttt{emchain} on the raw MOS-2 event list from each observation.  The ESAS tool \texttt{espfilt} was then run on the resulting event lists, to identify periods of soft proton flaring in each CCD and times when individual CCDs were operating in ``anomalous'' states with elevated backgrounds.  One output of \texttt{espfilt} is a GTI file that can be applied to remove the flagged intervals.  The ESAS tool \texttt{cheese} was then run in order to detect and excise point sources from the data.  Given the high column density expected along the line of sight to G351.9$-$0.9, we ran this tool in soft and hard bands; bright point sources were primarily detected in the soft band, indicating that they lie in the foreground rather than within or behind G351.9$-$0.9.  The ESAS tools \texttt{mosspectra} and \texttt{mosback} were run on each event list to create spectra and images of the quiescent particle background.  These tools rely on calibration files that are essentially archival observations that were taken with the filter wheel in the ``closed'' position.  It must be noted that the quiescent particle background is not evenly distributed across the face of the detector; rather, it is diffuse but concentrated close to the optical axis, and can easily be mistaken for diffuse emission from the targeted source.  This is particularly true for the MOS cameras, since the quiescent particle background spectra include strong Al K$\alpha$ and Si~K$\alpha$ lines at 1.49~keV and 1.75~keV, respectively.  Even using the ESAS tools, these lines cannot be entirely removed from resultant images and spectra, and they are unfortunately similar to lines expected in clusters of galaxies and supernova remnants.  Again owing to the high column density along the line of sight to G351.9$-$0.9, we did not correct for solar-wind charge exchange emission, since this contributes most at very low energy and the remnant should only be evident at higher energy.

Finally, the ESAS tool \texttt{combimage} was used to combine the products created from each observation, to create summed, filtered, fluxed images, corrected for exposure, soft proton flaring, the quiescent particle background, and point sources.  These images are shown in Figure \ref{fig:snr_xray}.  The left panel shows the combined MOS-2 image in the 2.0--7.0 keV band, with a representative set of contours from the radio image of G351.9$-$0.9 marking its location.  In this band, there is clearly no excess X-ray flux in the region of the remnant.  For comparison, the right panel shows the image in the 0.3--1.0~keV band; again, there is no indication of enhanced flux in the region of the remnant.  We therefore conclude that the remnant was not detected in X-rays, even in the deep combined exposure.

In order to estimate an upper limit on the X-ray flux from G351.9$-$0.9, we extracted spectra from each observation, using standard tools and an elliptical extraction region containing the remnant.  The spectra were then added using the FTOOLS \texttt{addascaspec} and \texttt{addrmf}, and binned using the ``optimal binning'' algorithm of \citet{kaastra2016}.  The summed spectrum was fit over the 0.3-10.0 keV band, minimizing a $\chi^{2}$ statistic.  We find that a model consisting of two thermal plasmas (via ``mekal'' in XSPEC) with temperatures of $kT = 0.88~keV$ and $kT = 0.45$~keV, two Gaussian lines centered at $E=1.49$~keV and $E=1.75$~keV, and a broken power-law ($\Gamma_{1} = 0.46$, $E_{br} = 5.7$~keV, $\Gamma_{2} = -3.5$) can describe the flux.  This model does not achieve an acceptable fit ($\chi^{2} = 584$ for $\nu = 131$ degrees of freedom), but it does capture the contributions from solar-wind charge exchange, the quiescent particle background, and residual soft proton flaring.  This model measures a flux of $F = 2.2\times 10^{-11}~{\rm erg}~{\rm cm}^{-2}~{\rm s}^{-1}$ (0.3-10.0 keV), and we therefore set this as an upper limit on the X-ray flux of G351.9$-$0.9.  These results were checked using the Swift snapshot observation of G351.9$-$0.9.  The shape of the astrophysical background is similar, but the detector backgrounds differ.  The background observed in the much deeper XMM-Newton exposures offers much tighter constraints despite its complexity.

The interaction between the neutron star X-ray binary Cir X-1 and its surrounding SNR makes its association clear \citep{heinz2013}.  As a final check, then, we examined the Chandra OBSID 22289 for evidence of interaction between Swift J1728.9$-$3613 and its local environment.  Using the CIAO tool \texttt{specextract}, we extracted radial profiles and spectra using 75 annuli generated between 0--30'' for the source and between 35--40'' for the background, after correcting for the ACIS readout streak.  These resulting spectra were fit with a simple model, \texttt{tbabs$\times$powerlaw} model.  This model was input into the ChaRT ray tracing simulation. The resulting rays were input into MARX, where we simulated an observation with the same exposure time, with the ray tracing file output by ChaRT.  Initial fits to the spectrum of Swift~J1728.9$-$3613 in this observation are suggestive of moderate photon pile-up, manifested in a hard power-law index and potentially flattened radial profile.  The source flux that was used to generate MARX simulations was based on fits to the spectrum of Swift~J1728.9$-$3613 including the \texttt{pileup} model in XSPEC \citep{davis2001}, which returned a slightly higher flux than models without this component.  The emission profiles that result are broadly consistent, and indicate that any extended emission in the regions immediately surrounding Swift J1728.9$-$3613 is due to photon pile-up.  There is no evidence of extension, especially not asymmetric, jet-like extension, that would indicate an interaction with the remnant.

\begin{figure}[!t]
    \hspace{-0.25in}
    \includegraphics[width=0.55\textwidth]{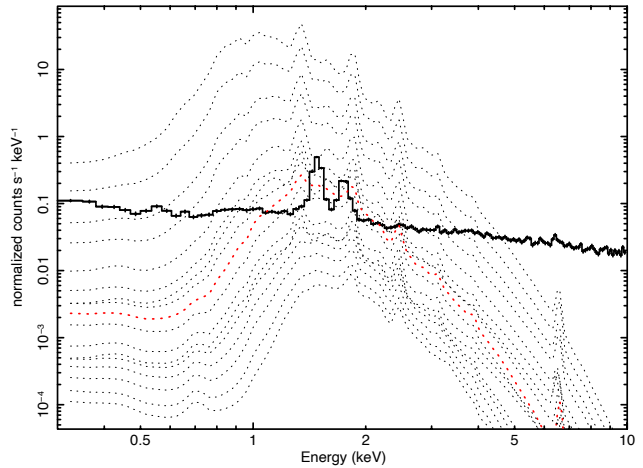}
    \caption{Anticipated absorbed Sedov emission from G351.9$-$0.9 versus instrumental and astrophysical backgrounds.  The data (in solid black) represents the observed combination of instrumental and astrophysical backgrounds observed with XMM-Newton.  The models (dashed lines) show the predicted absorbed Sedov emission from G351.9$-$0.9 as a function of distance in 0.5~kpc steps, from 4~kpc (top) to 12~kpc (bottom).  The model in red depicts the absorbed emission from $d=7.5$~kpc; at this distance, the expected emission is marginally detectable, though it peaks close to the instrumental Al~K$\alpha$ emission line at 1.49~keV.  We therefore set this as a lower limit on the distance to G351.9$-$0.9.  Please see the text for additional details.}
    \label{fig:sims}
\end{figure}

\subsection{The Distance to G351.9-0.9}\label{sec:snr_dist}

Observations have established a relationship between the radio surface brightness of a SNR, $\Sigma$, and its diameter, $D$.  The diameter and angular size of a given remnant can then be used to determine its distance.  Our MeerKAT observtions yield a total flux density of 1.6--1.8~Jy; assuming major and minor axes of 12' and 9', this translates into a surface brightness of $\Sigma_{1.28 \ GHz} = 2.23 - 2.51 \times \ 10^{-21} \rm W \ m^{-2} \ Hz^{-1} \ sr^{-1} $. Recent statistical treatments of catalogs by \citet{pavlovic2014} and \citet{vukotic2019} have derived improved $\Sigma-D$ relationships.  \citet{pavlovic2014}  employed a probability density function (PDF) to derive the distance to G351.9$-$0.9, which has the advantage of retaining more information about the calibration sample than a simple linear fit.  Mean, median, and mode distances of 8.5~kpc, 8.5~kpc, and 7.9~kpc were derived using the PDF method, broadly consistent with a value of 9.0~kpc derived using an orthogonal fitting method.  For a derived diameter of $D = 27.2$~pc, the results of \citet{xu2005} give a remnant age of $\tau = 15,190$~years.  It must be noted that \citet{pavlovic2014}  do not quote errors for their diameter and distance estimates, and the differences between the methods may be less than the uncertainty in each.   \citet{vukotic2019} examine a number of selection effects and potential sub-samples, and derive a $\Sigma-D$ relationship for each.  Their values are consistent with those derived by \citet{pavlovic2014} , but the errors allow for distances as large as 30~kpc.  

The fact that G351.9$-$0.9 is likely in the Sedov phase, and therefore likely dominated by relatively soft X-ray emission, means that the line of sight column density and observed backgrounds could prevent its detection if it lies at a sufficiently large distance.  This can be translated into an effective lower limit on the distance to G351.9$-$0.9 that is more constraining than the $\Sigma-D$ relationship.   With the assumption that the total explosion energy is $10^{51}~{\rm erg}$, the Sedov solutions at a given distance are completely determined by the fixed angular size of the remnant and the density of the local ISM.  We previously developed a characterization of the ISM density as a function of distance along this line of sight; this also determines the obscuring column density at a given distance.  A set of distinct, independent Sedov spectra, each determined by the unique conditions at a given point along the line of sight, can therefore be calculated.

The XSPEC implementation of the Sedov model \citep{borkowski2001} captures the emission in terms of the mean shock temperature, the electron temperature behind the shock, the ionization age of the remnant ($\tau_{ion} = n\times\tau$, where $n$ is the local ISM number density and $\tau$ is the remnant age), the metal abundances (relative to solar), and a flux normalization that is simply the emission measure diluted by distance.  The age of the remnant can be determined via the equation $r = 12.9 t_{4}^{2/5} (\epsilon/n)^{1/5}$~pc (where $r$ is the effective radius at a given distance, $\epsilon$ is the explosion energy in units of $10^{51}~{\rm erg}$, and $t_{4}$ is the remnant age in $10^{4}$~years; see \citealt{cioffi1988}); the ionization age is then also determined.  The mean shock temperature is given by $T_{\rm mean} = 3\times 10^{6} t_{4}^{-6/5} (\epsilon/n)^{2/5}$~K \citep{cioffi1988}, and it is related to the electron and ion temperatures via $T_{mean} = \frac{T_{e}+T_{i}}{2} = 1.35\times 10^{5} v_{100}^{2}$ and $\frac{T_{e}}{T_{i}} = \frac{4^{2}}{v_{100}^{2}}$ (where $v_{100}$ is the shock speed in units of $10^{2}~{\rm km}~{\rm s}^{-1}$; see \citealt{ghavamian2007}).

Within XSPEC, then, we constructed a set of absorbed Sedov spectra corresponding to 0.5~kpc intervals between 4.0--12.0~kpc. Each spectrum was simulated using the ``fakeit'' command, utilizing the observed MOS-2 background described above and the corresponding MOS-2 response files.  In each case, the simulated source and background exposure times were set to be equivalent to the total XMM-Newton exposure on G351.9$-$0.9 (184~ks).

Figure \ref{fig:sims} shows the results of these simulations.  The absorbed spectra are shown as a sequence of dashed lines.  Their emission is very soft, owing to relatively low electron temperatures.  Models corresponding to smaller distances and less-absorbed Sedov emission spectra lie above the observed MOS-2 background (the data plotted as connected black points).  In contrast, models corresponding to large distances and correspondingly higher levels of obscuration lie below the MOS-2 background, and would not have been detected in the combined 184~ks XMM-Newton exposure.  The models that just exceed the background only do so in the 1.1-1.4~keV band, with the column density dominating at low energy and the instrumental line dominating at 1.49~keV.  The model in red corresponds to the expected absorption and emission at a distance of $d=7.5$~kpc.  In 184~ks, this model would nominally produce 15,000 counts above background.  On this basis, we suggest that the remnant would nominally be detected at 7.5~kpc and smaller distances; so $d \geq 7.5$~kpc. At 7.5~kpc, the remnant would be $\sim$20-25 pc in diameter, which is consistent with our interpretation of the remnant's morphology.

Although a slightly larger distance and higher column density might yield a weaker detection, it is reasonable to be conservative in this circumstance.  The derived distance limit is based on an assumed explosion energy, and solar abundances.  It also hinges on our characterization of the MOS-2 background, and therefore the efficacy of our efforts to characterize the soft proton flaring and the accuracy of quiescent particle background calibrations.  Moreover, the accuracy with which emission can be separated from an instrumental background -- particularly a background with strong lines -- is necessarily over-estimated in simulated data.  In practice, excluding a broader range of energies around the instrumental lines would be pragmatic.

\section{Probability of Coincidental Overlap} \label{sec:prob}

\begin{figure}[h!]
    \centering
    \includegraphics[width=0.48\textwidth]{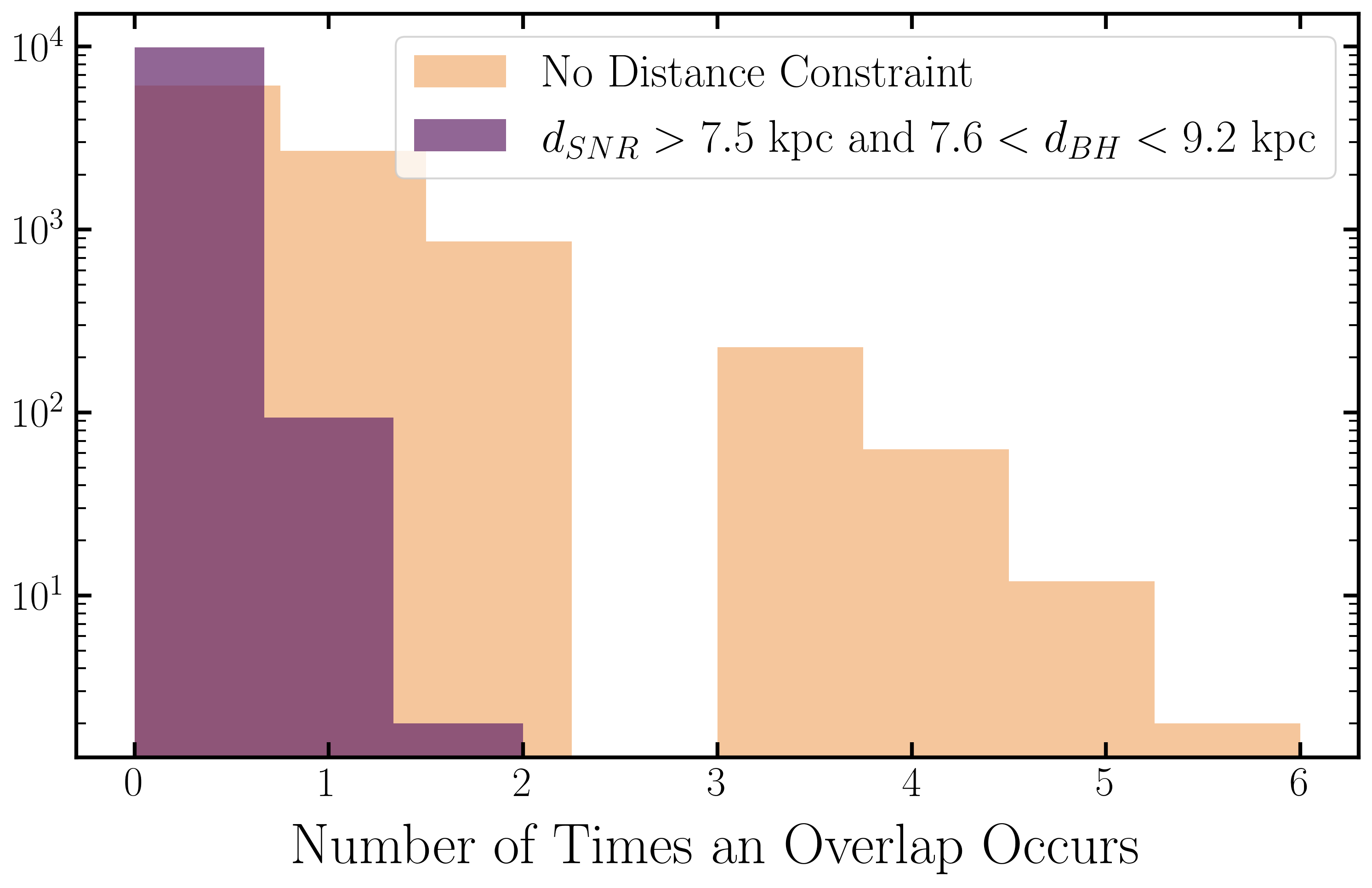}
    \caption{We show a histogram of the number of overlaps recorded from our simulation of 10,000 MW-like galaxies. Orange bars denote runs with no distance constraints, while purple bars refer to runs where we imposed the distance constraints we derive in this work. Note that the y-scale is in logarithmic units. It is clear that a distance constraint significantly decreases the number of times an overlap occurs between a BH and SNR. On average, there are 0.007 distance-constrained overlaps in each simulated galaxy. }
    \label{fig:overlap}
\end{figure}

To accurately determine the probability of a chance alignment, we use Milky Way supernova remnant (SNR) and black hole (BH) \citep{blackcat} catalogs to get positions of all known SNRs and BHs in the Milky Way. Please see Appendix \ref{sec:ap_prob} for details.  We created distributions for galactic latitudes, longitudes, distances, and angular sizes and sampled from MW catalogs independently to ``simulate'' a galaxy. We sampled all parameters separately (e.g., galactic latitudes for SNRs and then galactic latitudes for BHs) and then used rejection sampling to draw a new set of values which were combined with other new parameter values to create a simulated galaxy. 
 
For example, we would take the MW distribution of SNRs Galactic latitudes, and use rejection sampling to draw a new distribution of SNR latitudes. Then, we do the same with SNR distances, and all other parameters (other than SNR angular distance and longitude, which were sampled together), and combine all the new values to ``create'' new SNRs and BHs that would have new distances, longitudes, and latitudes. When looking at these parameters separately, e.g., new SNR distances, new SNR latitudes, etc., they would have probability distributions that are similar to the distribution we see in our galaxy. The end result was galaxies with populations of BHs and SNRs with l, b, distance, and size values that had probability distributions that mimic the probability distributions of these parameters observed in the Milky Way. 
In our simulations, an overlap occurred if a BH is found  within the central half of the SNR, i.e. the BH had coordinates within half of the angular height and width of the SNR on the sky, and it was within 90$^\circ$ of the GC (details in Appendix \ref{sec:ap_prob}). We simulated 10,000 galaxies, and recorded the number of overlaps.  We can constrain the radial distance of the simulated BH and SNR with the results of this work, that $d_{SNR} > 7.5$ kpc (see Section \ref{sec:snr_dist}) and $d_{BH} = 8.4 \text{ kpc} \pm 0.8 $ kpc (see Section \ref{sec:bh_dist}).  When we impose this distance constraint, the number of chance overlaps decreases significantly, as seen in Figure \ref{fig:overlap}. Here, we show a histogram of the number of overlaps, with (purple) and without (orange) our distance constraint (note that the y-axis is in log-scale).  The average number of overlaps per galaxy with constrained radial distance is 0.007, while without a distance constraint this increases to an average of 0.52 overlaps per galaxy. In other words, a given MW-type galaxy has a 0.7\% probability of having a BH overlap with the central half of a SNR within 90$^\circ$ of the GC and with our distance constraints. 

There are other ways of representing the histogram in Figure \ref{fig:overlap}. If we divide the number of overlaps by the number of BHs in every simulated galaxy, we can calculate the probability of a given BH being located at a position covered by the central half of a SNR (details in Appendix \ref{sec:ap_prob}).  The probability of finding a BH coincident with a SNR's central half of $P < 6.667\%$ without a distance constraint and $P < 1.695\%$ with a distance constraint, where we report the 3$\sigma$ upper limit from our simulations. 

The number of known supernova remnants is limited by current observing capabilities, but new telescopes such as MeerKAT are leading the way in the search for supernova remnants. As we observe lower flux density supernova remnants, the number of known supernova remnants will increase. Linearly extending the current distribution \citep{green2019} of SNRs down to a radio flux density of 0.001 Jy results in a two-fold increase in the number of SNRs. Increasing the number of SNRs by a factor of 2 results in an increase in the average number of overlaps to $P_{avg} \approx 2.5\%$.  With these results, we find it unlikely but not impossible that the overlap of Swift J1728.9$-$313 and G351.9$-$0.9 is coincidental.

\section{Discussion}

We have analyzed X-ray, radio, and IR observations of the transient X-ray binary Swift~J1728.9$-$3613, which lies within the interior of the known supernova remnant G351.9$-$0.9 in the plane of the sky.  Our analysis confirms and extends independent examinations of Swift J1728.9$-$3613 that conclude that the X-ray binary system harbors a black hole primary \citep{enoto2019,saha2022}.  We identified an IR counterpart in archival observations that is spatially coincident with the best X-ray position of Swift~J1728.9$-$3613, and underwent a contemporaneous outburst.  The quiescent colors of this counterpart are broadly consistent with a relatively massive A or B-type star, likely indicating that Swift~J1728.9$-$3613 is an intermediate-mass X-ray binary (IMXB) or high-mass X-ray binary (HMXB).  Utilizing the measured distribution of gas along the line of sight to Swift J1728.9$-$3613 and G351.9$-$0.9, we derived a distance to the X-ray binary system of $d = 8.4\pm 0.8$~kpc, and a lower limit on the distance to the SNR of $d \geq 7.5$~kpc.  We ran extensive simulations to determine the probability of a coincidental overlap between Swift J1728.9$-$3613 and G351.9$-$0.9, based on the known numbers and positions of black holes and supernova remnants in the Milky Way, the distribution of remnant sizes, and these distance constraints.  We find that the probability of coincidental overlap between a black hole and a remnant is strictly $<1.7\%$.  This suggests that the two are likely to be physically connected.  

If the same core-collapse supernova created the black hole in Swift~J1728.9$-$3613 and G351.9$-$0.9, the most fundamental implication is that not all black holes are ``born in the dark.''  Rather, distinct formation channels may exist for black holes.  This would represent observational support for numerical simulations that predict a subset of core-collapse supernovae can leave black holes and remnants \citep{ugliano2012}.  It is worth repeating here that SS~433 in W50, MAXI J1535$-$571, and W49B may also represent such cases \citep{blundell2008, kubota2010, cherepashchuk2018, maxted2020, lopez2013}.  Unlike SS 433 and W49B, however, X-rays clearly detect a compact object and strongly favor a black hole primary in Swift~J1728.9$-$3613, and for these reasons it may represent the strongest case for a distinct black hole formation channel.  Another fundamental implication is that the black hole in Swift~J1728.9$-$3613 must be very young, and its high spin ($a = 0.86\pm 0.02$) must be set at birth rather than through accretion.  This is the subject of a companion paper; please see Draghis et al.\ (2023).

Binary evolution models generally predict that much longer time scales are required for accretion to commence, than is permitted by the potential youth of Swift~J1728.9$-$3613 and G351.9$-$0.9.  For example, \citet{podsiadlowski2003} suggests that $\tau \simeq 1$~Gyr is required for mass transfer rates of $\dot{M} \simeq 10^{-9} M_\odot$ per year.  While \citet{podsiadlowski2003} accurately predicts the mass of the companion and the spin of the black hole in GRS 1915$+$105, the same considerations predict a companion that is less massive than the black hole in Cygnus X-1.  In contrast, observations find that the companion is about twice as massive as the black hole \citep{miller-jones2021}.  Moreover, Cir X-1 is transient neutron star X-ray binary in a remnant with an age of $\tau \leq 4600$~years \citep{heinz2013}, and it also has an A- or B-type companion star.  It may be the case that stronger binary evolution is possible before supernova events in some instances, leading to close, circularized binaries.  

The binary evolution that is required if Swift J1728.9$-$3613 is physically associated with G351.9$-$0.9 is broadly consistent with that described by \citet{maxted2020}.  Motivated by the possible association of MAXI~J1535$-$571 with G323.7$-1$.0, \citet{maxted2020} employed the population synthesis code ``StarTrack'' \citep{StarTrack}.  This code has been used to model X-ray binary populations in other galaxies; in NGC 1569, it predicts a luminosity function that is within a factor of two of the observed luminosity function \citep{Belczynski}.  StarTrack also made key predictions connecting short gamma-ray bursts to mergers of BH-BH/BH-NS/NS-NS binaries that were confirmed by LIGO in 2017 \citep{ligo2017}.  Using StarTrack, \citet{maxted2020} found three different binary evolution paths that could lead to a low-mass X-ray binary or IMXB in a SNR.  The evolutionary paths start with a $\sim 25~M_\odot$ primary and $\sim3~M_\odot$ companion, and result in a system with a 5-11~$M_\odot$ black hole and a companion with a mass of 2.6-4.5 $M_\odot$ (an A or B type star) in a very close orbit, such that mass transfer begins approximately $\tau \simeq 24,000$~years after the supernova.  Within the bounds of the data that are currently available, these predictions broadly match Swift J1728.9$-$3613 and G351.9$-$0.9.

New numerical simulations are needed to understand the relative rates of black hole production through different channels, and understand binary evolution mechanisms that allow for the formation of a highly spinning black hole (see e.g., \citealt{qin2019}).  While the absolute number of supernova remnants with evidence of black holes within the Milky Way is likely to be small, prompt observations of supernovae in nearby galaxies are beginning to reveal aspects of compact object formation \citep{pasham2021}, and may soon provide another window into black hole formation.   In the longer run, the launch of the Athena mission will deliver sensitive, high-resolution spectra of supernova remnants in nearby galaxies; comparisons to remnants hosting black holes in the Milky Way will help to understand black hole formation.

Especially given the broad implications of the potential association between Swift J1728.9$-$3613 and G351.9$-$0.9, it is vital to critically examine the evidence that Swift J1728.9$-$3613 harbors a black hole, and the likelihood that it is physically associated with G351.9$-$0.9.  It is equally vital to examine alternative interpretations of the data.  The rest of this section examines caveats of our analysis, and discusses possible alternatives.

The 5.5~Hz QPOs observed in the NICER power spectra of Swift J1728.9$-$3613 are consistent with Type B QPOs that are typically found in the soft-intermediate state of black hole X-ray binaries \citep{belloni2012}.  However, observations and theory do not dictate that these oscillations cannot be found in neutron stars.  The evolution of the X-ray colors observed in NICER observations may be a stronger indicator of a black hole primary in Swift J1728.9$-$3613.  The source clearly populates a region that is exclusive to black hole systems in a prior broad analysis, but two caveats must be attached to our color analysis: (1) Swift~J1728.9$-$3613 has a higher column density than the ones upon which this diagnostic is based \citep{done2003}, and (2) the NICER data that we employed does not extend above 10~keV.  In principle, both concerns can be mitigated in the limit of very sensitive data; moreover, we examined three different NICER background estimation methods in order to check the validity of extrapolating more models to higher energies.  A more exhaustive analysis of X-ray binary colors than undertaken by \citet{done2003} and/or the discovery of new transient neutron star X-ray binaries may reveal a source or sources that populate the ``black holes only'' region of the color-color diagram, invalidating this diagnostic.
The value of the spin parameter measured with NuSTAR, $a = 0.86\pm 0.02$ (Draghis et al.\ 2023), may represent the most rigorous and quantitative evidence that Swift J1728.9$-$3613 harbors a black hole.  The error on the spin parameter excludes the theoretical maximum spin for a neutron star at the $5\sigma$ level of confidence.   While the uncertainties given by Draghis et al.\ 2023 are simply statistical, they perform a number of tests in order to assess any systematic errors of the measurement, and no consideration points to a spin parameter that is consistent with the range feasible for neutron stars.  All measures of black hole spin that utilize the disk depend upon the assumption that it extends to the ISCO; NuSTAR observed Swift J1728.9$-$3613 at an Eddington fraction where this assumption is likely valid (\citealt{draghis2022}; also see \citealt{shafee2008,reynolds2008}).  Although the statistical mode of the black hole spin distributions derived using relativistic disk reflection and disk continuum spectroscopy are formally consistent ($a_{cont} = 0.908^{+0.086}_{-0.558}$, $a_{refl} = 0.968^{+0.028}_{-0.240}$, \citealt{draghis2022}), the continuum-derived distribution includes some low spin values, and the two methods give statistically inconsistent results in a few cases.  It is currently not possible to test the relativistic reflection results using the disk continuum of Swift J1728.9$-$3613, since the black hole mass cannot be determined using optical techniques.
We have also conducted systematic searches for coherent pulsations and Type I X-ray bursts in Swift J1728.9$-$3613, since these phenomena clearly indicate that a given X-ray binary harbors a neutron star primary.  No pulsations are detected in Swift J1728.9$-$3613, and the upper limits we derive are comparable to the pulsation amplitudes detected in a number of X-ray binaries.  However, it is worth recalling that a coherent pulsation at $\nu = 550.27$~Hz was found in just 150~s of data in the X-ray binary Aql X-1, amid a total exposure time of 1.3 Ms \citep{casella2008}.  Similarly, we do not find any Type I X-ray bursts in an extensive search of the light curves obtained from Swift J1728.9$-$3613, but our observing campaign could have missed a burst from a source like Cir X-1 that has a long and/or erratic recurrence time.  \citep{tennant1986, linares2010}.  The null results of our searches do not conclusively rule out a neutron star primary, and more exhaustive observing campaigns during a future outburst of Swift J1728.9$-$3613 may be able to detect pulsations or a Type I burst. 

Our simulations indicate that a false association between Swift~J1728.9$-$3613 and G351.9$-$0.9 is improbable, but not impossible.  We have attempted to anticipate the effect of a higher number of Galactic supernova remnants that may be discovered in new radio surveys; the probability of a chance association remains low, but it does grow accordingly.  The radio image of G351.9$-$0.9 is fully consistent with a supernova remnant in the Sedov phase, and this is bolstered by its surface brightness.  However, it is not possible to completely exclude the possibility that the structure is actually a superbubble resulting from multiple supernovae and/or massive stellar winds.  Thermal X-ray emission is seen in many superbubbles and some cases appear to include non-thermal X-ray emission \citep[e.g.][]{sasaki2022}; if G351.9$-$0.9 is a superbubble, its non-detection with XMM-Newton indicates that it likely does not lie in the foreground, but instead in the vicinity or background of Swift J1728.9$-$3613.

\begin{acknowledgements}
We would like to thank the directors and teams at MeerKAT, XMM-Newton, NuSTAR, Swift, Chandra, and NICER.  NICER work at NRL is supported by NASA.  The MeerKAT telescope is operated by the South African Radio Astronomy Observatory, which is a facility of the National Research Foundation, an agency of the Department of Science and Innovation. AZ is supported by NASA under award number 80GSFC21M0002.  We acknowledge helpful conversations with Teruaki Enoto, Duncan Galloway, Marko Pavlovic, Steve Snowden, and Branislav Vukotic. The authors would like to thank the anonymous referees for their comments and suggestions, which greatly improved the quality of our manuscript.
\end{acknowledgements}

\section{Software \& Data}

\software{CIAO (v4.13;  \citet{ciao}), SAS (v1.3; \citet{xmmsas}) v20.0.0 with a development version of ESAS based on \citet{snowden2008} that will be released with SAS v21 , HEAsoft (v6.28; \citet{heasoft}), nibackgen3C50 \citep{remillard3c50}, nicer-background (v0.4.t1.200e; Zoghbi et al. in prep; \url{https://github.com/zoghbi-a/nicer-background}) stingray \citep{stingray_a,stingray_b}, XSPEC \citep{xspec}, oxkat \citep{oxkat2020}, CASA \citep{casa}, CARTA \citep{carta}, PRESTO \citep{presto}, Scipy \citep{scipy}}

The data and code used in our probability of overlap analysis can be found at this link \url{https://zenodo.org/badge/latestdoi/568961617}.

\startlongtable
\begin{deluxetable*}{c|c|c|c|c}
\tablehead{
\colhead{Telescope} & \colhead{OBSID} & \colhead{Start Time (MJD)} & \colhead{Exposure Time (ks)} & \colhead{Target} 
}
\startdata
NICER & 1200550101 & 58512 & 1.3 & Swift J1728.9$-$3613\\
NICER & 1200550102 & 58513 & 11.7 & Swift J1728.9$-$3613\\ 
NICER & 1200550103 & 58514 & 14.1 & Swift J1728.9$-$3613\\ 
NICER & 1200550104 & 58515 & 10.6 & Swift J1728.9$-$3613\\ 
NICER & 1200550105 & 58516 & 8.2 & Swift J1728.9$-$3613\\ 
NICER & 1200550106 & 58517 & 19.1 & Swift J1728.9$-$3613\\ 
NICER & 1200550107 & 58518 & 10.7 & Swift J1728.9$-$3613\\ 
NICER & 1200550108 & 58519 & 7.4 & Swift J1728.9$-$3613\\ 
NICER & 1200550109 & 58520 & 7.4 & Swift J1728.9$-$3613\\ 
NICER & 1200550110 & 58521 & 3.4 & Swift J1728.9$-$3613\\ 
NICER & 1200550111 & 58522 & 1.5 & Swift J1728.9$-$3613\\ 
NICER & 1200550112 & 58523 & 1.4 & Swift J1728.9$-$3613\\ 
NICER & 1200550113 & 58524 & 5.0 & Swift J1728.9$-$3613\\ 
NICER & 1200550114 & 58525 & 4.2 & Swift J1728.9$-$3613\\ 
NICER & 1200550115 & 58526 & 8.0 & Swift J1728.9$-$3613\\ 
NICER & 1200550116 & 58527 & 4.5 & Swift J1728.9$-$3613\\ 
NICER & 1200550117 & 58528 & 1.1 & Swift J1728.9$-$3613\\ 
NICER & 1200550118 & 58529 & 1.9 & Swift J1728.9$-$3613\\ 
NICER & 1200550119 & 58530 & 3.2 & Swift J1728.9$-$3613\\ 
NICER & 1200550120 & 58531 & 0.8 & Swift J1728.9$-$3613\\ 
NICER & 1200550121 & 58532 & 0.9 & Swift J1728.9$-$3613\\ 
NICER & 1200550122 & 58533 & 0.7 & Swift J1728.9$-$3613\\ 
NICER & 1200550123 & 58534 & 1.2 & Swift J1728.9$-$3613\\ 
NICER & 1200550124 & 58535 & 1.4 & Swift J1728.9$-$3613\\ 
NICER & 1200550125 & 58539 & 3.5 & Swift J1728.9$-$3613\\ 
NICER & 1200550126 & 58540 & 2.3 & Swift J1728.9$-$3613\\ 
NICER & 2200550104 & 58560 & 1.2 & Swift J1728.9$-$3613\\ 
NICER & 2200550105 & 58561 & 1.9 & Swift J1728.9$-$3613\\ 
NICER & 2200550108 & 58571 & 1.7 & Swift J1728.9$-$3613\\ 
NICER & 2200550110 & 58572 & 2.5 & Swift J1728.9$-$3613\\ 
NICER & 2200550111 & 58573 & 1.8 & Swift J1728.9$-$3613\\ 
NICER & 2200550112 & 58574 & 3.1 & Swift J1728.9$-$3613\\ 
NICER & 2200550113 & 58575 & 1.8 & Swift J1728.9$-$3613\\ 
NICER & 2200550115 & 58585 & 2.3 & Swift J1728.9$-$3613\\ 
NICER & 2200550117 & 58586 & 2.4 & Swift J1728.9$-$3613\\ 
NICER & 2200550120 & 58590 & 1.4 & Swift J1728.9$-$3613\\ 
NICER & 2200550122 & 58593 & 1.4 & Swift J1728.9$-$3613\\ 
NICER & 2200550123 & 58594 & 1.1 & Swift J1728.9$-$3613\\ 
NICER & 2200550124 & 58595 & 1.6 & Swift J1728.9$-$3613\\ 
NICER & 2200550130 & 58612 & 2.1 & Swift J1728.9$-$3613\\ 
NICER & 2200550133 & 58636 & 1.5 & Swift J1728.9$-$3613\\ 
NICER & 2200550137 & 58657 & 1.8 & Swift J1728.9$-$3613\\ 
NICER & 2200550138 & 58658 & 1.5 & Swift J1728.9$-$3613\\ 
NICER & 2200550139 & 58660 & 1.5 & Swift J1728.9$-$3613\\ 
NICER & 2200550140 & 58660 & 1.8 & Swift J1728.9$-$3613\\
NICER & 2200550142 & 58662 & 2.0 & Swift J1728.9$-$3613\\
NICER & 2200550145 & 58674 & 1.4 & Swift J1728.9$-$3613\\
NICER & 2200550147 & 58674 & 2.3 & Swift J1728.9$-$3613\\
NICER & 2200550151 & 58697 & 1.1 & Swift J1728.9$-$3613\\
NICER & 2200550152 & 58697 & 2.7 & Swift J1728.9$-$3613\\
Chandra & 21213 & 58526 & 5.1 & Swift J1728.9$-$3613\\ 
Chandra & 22289 & 58692 & 30.1 & Swift J1728.9$-$3613\\ 
NuSTAR & 90501303002 & 58518 & 16.8 & Swift J1728.9$-$3613\\ 
Swift & 00011201001 & 58554 & 1.4 & Swift J1728.9$-$3613\\ 
Swift & 00011201002 & 58694 & 2.1 & Swift J1728.9$-$3613\\ 
Swift & 00011201003 & 58697 & 3.0 & Swift J1728.9$-$3613\\ 
Swift & 00011201004 & 58746 & 2.9 & Swift J1728.9$-$3613\\ 
Swift & 00011201005 & 58748 & 1.8 & Swift J1728.9$-$3613\\ 
Swift & 00011201006 & 58750 & 1.9 & Swift J1728.9$-$3613\\ 
\hline
Swift & 00011201007 & 59844 & 1.5 & G351.9$-$0.9\\
XMM-Newton & 860140101 & 59116 & 18 & G351.9$-$0.9\\ 
XMM-Newton & 0901540101 & 59638 & 83 & G351.9$-$0.9\\ 
XMM-Newton & 0901540201 & 59642 & 83 & G351.9$-$0.9\\
\enddata
\vspace{1 mm}
\tablecomments{Telescope, OBSIDs, starting days, exposure times, and targets for the key observations in our analysis.  We have omitted observations the remaining Chandra observations for brevity and clarity.}
\label{tab:obsids}
\end{deluxetable*}

\begin{deluxetable*}{c|c|c|c|c|c}
\tablehead{
\colhead{Source Name} & \colhead{l (deg)} & \colhead{b (deg)} & \colhead{$N_H$ $(10^{22} cm^{-2})$} & \colhead{Distance from literature (kpc)}& \colhead{Distance from $N_H$ (kpc)}
}
\startdata
V404 Cyg & 73.1 & 2.1 & 6.4 \citep{kajava2020} & 2.4 $\pm$ 0.1 \citep{miller-jones2009} & 2.1 \\ 
GRS 1915+105 & 45 & -0.2 & 3.5 & 9 $\pm$ 3 \citep{chapuis2004} & 9.0 \\ 
SS 433 & 39.7 & -2.2 & 1.76 & 4.9 \citep{lockman2007} & 4.7 \\ 
GRO J1744-28 & 0 & 0.3 & 5 & 8 \citep{monkkonen2019} & 8.6 \\ 
XTE J1739-285 & 359.7 & 1.3 & 1.73 $\pm$ 0.01 & 4 $\substack{+4 \\ -2}$ \citep{bult2021} & 4.0 \\ 
IGR J19140+0951 & 44.3 & -0.5 & 6 & 5.2 \citep{hannikainen2007} & 5.0 \\ 
IGR J17544-2619 & 3.2 & -0.3 & 1.2 & 2 -- 4 \citep{pellizza2006} & 4.2 \\ 
SAX J1818.6-1703 & 16.9 & -1.3 & 1.17 \citep{kalberla2005} & 2.1 \citep{negueruela2006} & 3 \\ 
4U 1728-34 & 354.3 & -0.2 & 2.6 \citep{worpel2013} & 5.2 \citep{galloway2008} & 5.0 \\ 
GX 3+1 & 2.3 & 0.8 & 1.91 $\pm$ 0.05 \citep{rogantini2019} & 3 -- 7 \citep{kuulkers2000} & 4.2 \\ 
RX J1718.4-4029 & 347.3 & -1.7 & 3.1 $\pm$ 1.3 & 6.5 $\pm$ 0.5 \citep{cornelisse2004} & $5.9 \substack{+4.4 \\ -1.9} $\\ 
2S 1711-339 & 352.1 & 2.7 & 1.5 & $<$ 7.5 \citep{cornelisse2002} & 3.6 \\ 
XTE J1720-318 & 354.6 & 1 & 1.33 & 3 -- 10 \citep{chaty2006} & 3.2 \\
XTE J1723-376 & 350.2 & $-$0.9 & 1.2 \citep{kalberla2005} & 1.3 -- 10 \citep{galloway2008} & 3 \\ 
SAX J1713.4-4219 & 348.9 & 0.9 & 2 & 4.8 \citep{smith2012} & 4.5 \\ 
4U 1630-47 & 336.9 & 0.3 & 5.8 $\pm$ 0.3 \citep{tomsick2014} & 4.7 -- 11.5 \citep{kalemci2018} & 4.53 -- 9\\ 
G350.1-0.3 & 350.1 & -0.3 & 3.2 $\pm$ 0.1 & 9 $\pm$ 3 \citep{yasumi2014} & $6.5 \substack{+0.5 \\ -0.4}$ \\ 
\enddata
\vspace{1 mm}
\caption{This table contains references and values for Figure \ref{fig:NH_tests}, which shows comparisons between our distance estimation technique and published values. If no source is cited for the $N_H$ value, that value came from the same source used for the distance.}
\label{tab:refs}
\end{deluxetable*}

\begin{figure*}[!ht]
    \centering
    \includegraphics[width= 0.95 \textwidth]{ 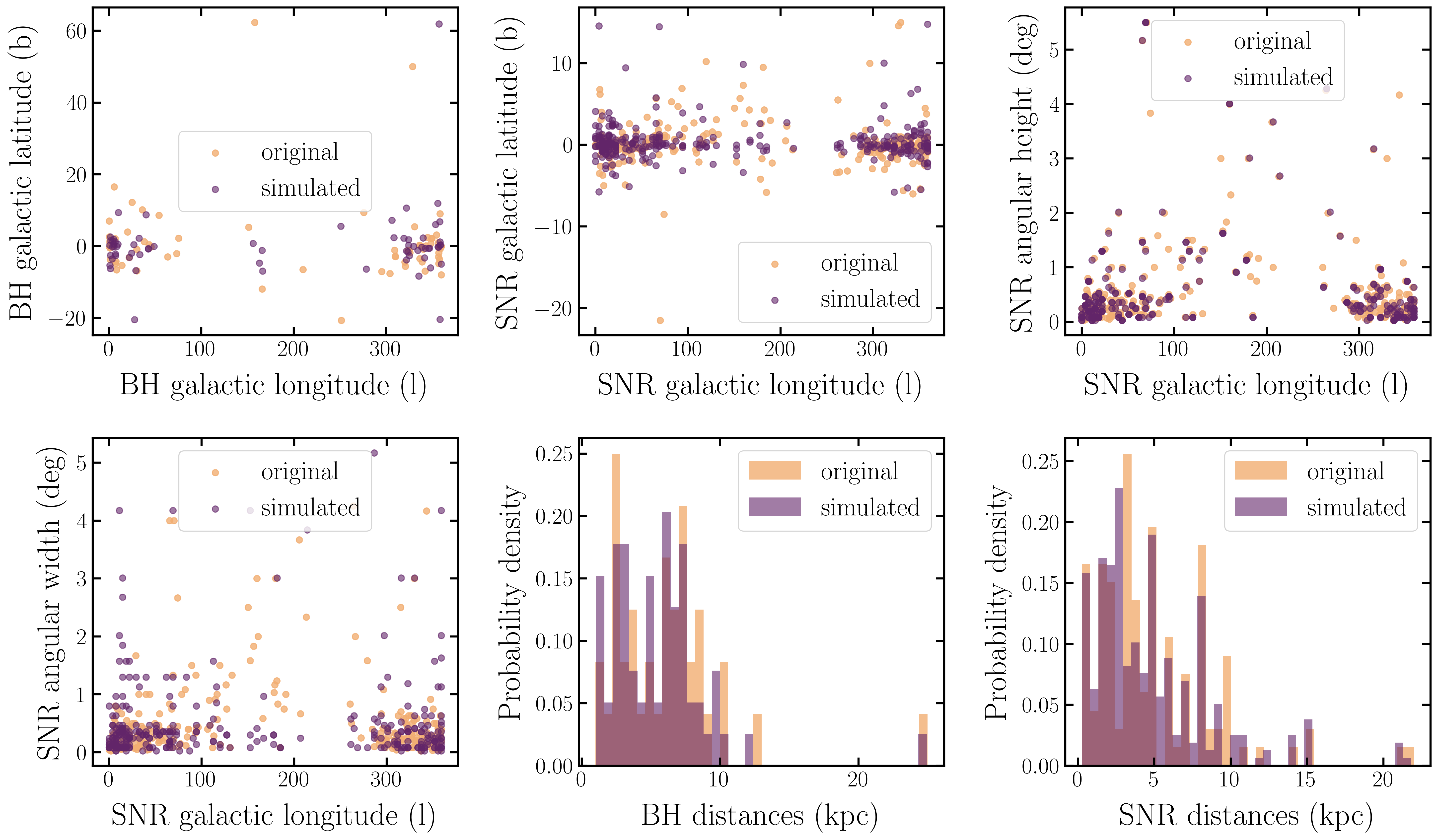}
    \caption{This figure shows how well the simulations in our probability analysis reproduce known Milky Way distributions for different populations. From top to bottom, left to right, the panels show plots of b vs. l (BHs), b vs. l (SNRs), angular height vs. l (SNRs), angular width vs. l (SNRs), and probability density histograms for BH distances and SNR distances. In all panels, the original data from MW catalogs is shown in orange, and the simulated populations are shown in purple.}
    \label{fig:simulation}
\end{figure*}

\begin{figure*}[h!]
    \centering
    \includegraphics[width=0.74\textwidth]{ 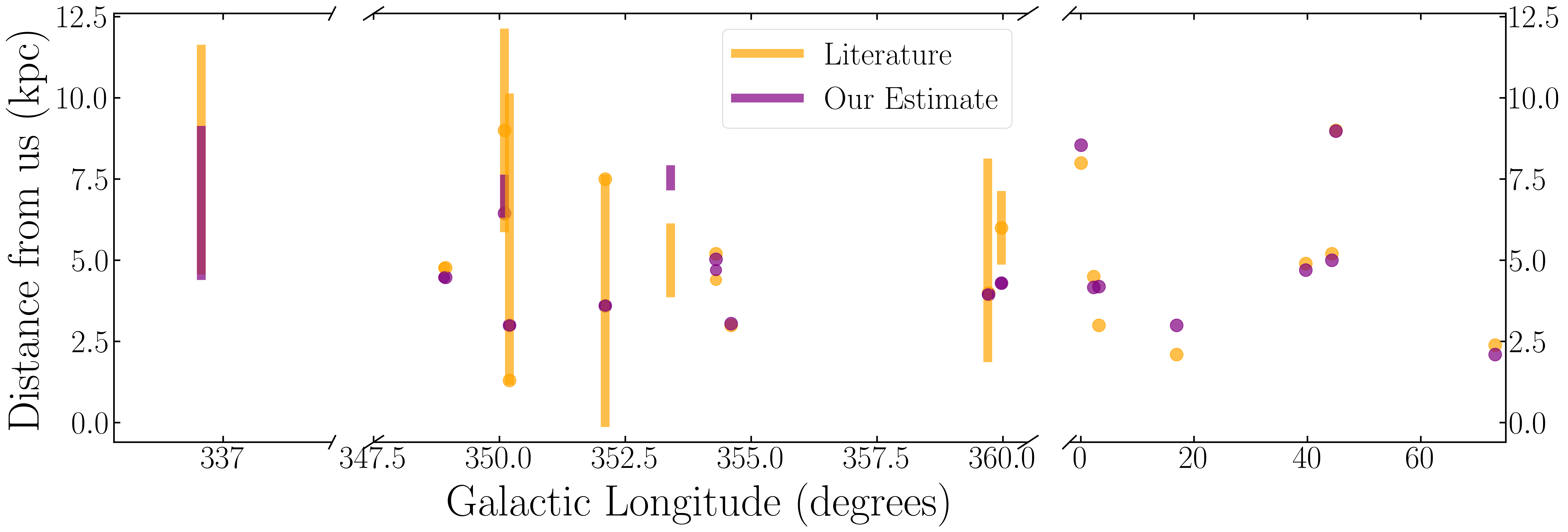}
    \caption{Comparisons of distances calculated using our technique (purple) of integrating the galactic $n_H$ distribution from column densities in the literature with published distances (orange).  Note the breaks in the galactic longitude in order to better visualize all the points. References are provided in Table \ref{tab:refs}. The red point corresponds to distances calculated using this method for the Chandra measurement of $N_H = 4.22 \pm 0.02 \times \rm 10^{22} cm^{-2}$ for Swift J1728.}
    \label{fig:NH_tests}
\end{figure*}

\appendix

\section{Distance Estimation Technique}\label{sec:ap_dist}

\noindent We tested our column density-based distance calculation on a set of well-known sources within 5 degrees of the Galactic plane in latitude and within $\sim$ 60 degrees of the Galactic Center in longitude.  These restrictions are appropriate for the hydrogen number density distributions that we used to create a continuous distribution \citep{kalberla2009, marasco2017}.  Our source sample draws heavily from the Galactic Center because structure within that region, molecular clouds, gas dynamics, superbubbles, and other factors are potentially important complications that could serve to indicate the method is not robust. \par

Table \ref{tab:refs} lists the position of each source, the reported column density along the line of sight, an independent distance estimate, and the distance estimate derived from the column density using our technique.  Figure \ref{fig:NH_tests} compares our technique with independent distance measurements from the literature.  The independent estimates are shown in orange, and the distances estimated using our technique are shown in purple.  The degree of correspondence is striking.  In many cases, there is formal agreement between the techniques; where the techniques differ, the fractional disagreement is less than 10\% in all cases but one, and in that instance (SAX J1818.6$-$1701) the absolute difference is less than 1~kpc ($d = 2.1$~kpc, \citet{negueruela2006}, versus $d = 3$~kpc using our technique).  \par

\section{Overlap Simulation}\label{sec:ap_prob}

We used Milky Way supernova remnant (SNR) \citep{green2019} and black hole (BH) \citep{blackcat} catalogs to get information for all known SNRs and BHs in the galaxy. We omitted SNR G351.9$-$0.9 (our remnant) from the SNR catalog. \par

For BHs, we used galactic latitudes, galactic longitudes, and distances from the catalogs. For the SNRs, we used galactic latitudes, galactic longitudes, distances, and angular sizes. For each parameter listed, a continuous probability distribution was approximated by inputting the bin centers and y-values on histograms into $\texttt{scipy.interpolate.interp1d}$. We sampled each probability distribution in order to obtain a new distribution with approximately the number of those objects in the galaxy using Monte Carlo rejection sampling. Rejection sampling is a technique where a grid of points is 'thrown' on a histogram of a given distribution, and all points below the curve of the histogram get accepted. A histogram of the new samples will return a probability distribution very similar to that of the initial data. \par
It was important to simulate galaxies that had approximately the same number of objects as the Milky Way: 294 SNRs and 62 BHs. In order to account for this, we ran our rejection sampling code on each parameter distribution until we reached an initial sample that would return, on average, the number of those objects in the Milky Way. We note that many objects did not have distance measurements, so we had to use more samples to simulate distance distributions, but we ensured that approximately 294 SNRs and 62 BHs were generated each time. The code used for this analysis is available at \url{https://zenodo.org/badge/latestdoi/568961617} with all relevant files. \par
The angular size of supernova remnants is closely tied to its galactic longitude (and not as closely to its galactic latitude in our work). Remnants with larger angular sizes are mostly observed towards the galactic anticenter. This is physically reasonable; SNRs toward the anti-center are closer to us than towards the Galactic Center, and will have larger angular sizes, and SNRs in the galactic centre could also be physically smaller because of the higher ambient gas density. This means that any coincidental overlaps are more likely to occur toward the Galactic anti-center, where there are more remnants that have large angular sizes. In order to account for the dependence between angular size and Galactic longitude, we sampled from a 2-D probability distribution, where 1-D rejection sampling was not sufficient. Here, we used the Walker alias method \citep{Walker1974} to randomly sample the distribution. Using the alias method samples directly from the initial values, rather than sampling from an initial probability distribution. \par
In Figure \ref{fig:simulation}, we show an example of a simulated galaxy's parameters when compared to our galaxy. The simulation technique used in our code very accurately samples from known values for SNRs, LMXBs, HMXBs, and BHs. Without sampling SNR l and size together, we wouldn't be able to recreate those distributions when creating our simulated galaxies. \par
To calculate the probability of overlap in a simulated galaxy, BH galactic l, b, and distance and SNR l, b, distance, and angular size, we iterated through each simulated supernova remnant and treated it like a box. If a simulated BH's galactic latitude and longitude placed it within half of the supernova remnant's angular height and width, and that supernova remnant was located within $90^\circ$ of the Galactic Center, it would count as an overlap. Our distance constraint required $d_{SNR} > 7.5$ kpc and $7.6 \text{kpc } < d_{BH} < 9.2$ kpc. \par
Our code simulates 10,000 Milky-Way like galaxies and returns the number of overlaps, the number of simulated SNRs, and the number of simulated BHs. There are several ways to convey these results. We can look at the average number of overlaps (the sum of the overlaps divided by the number of galaxies), the probability of an SNR being on top of a BH (number of overlaps divided by number of BHs), or the probability of a BH being inside a SNR (number of overlaps divided by number of SNRs). \par
With a distance constraint, the average number of overlaps tells us the likelihood that a MW-type galaxy would have a BH in the central half of the SNR where the SNR is at least 7.5 kpc from Earth and the BH is between 7.6 and 9.2 kpc from Earth. There are also other ways to statistically interpret our results. For a given BH, what is the probability of it overlapping with a SNR? Or for a given SNR, what is the probability of it appearing to contain a BH? \par
To demonstrate the difference between these two questions, we can introduce an analogy. First, imagine a wall with all known SNRs.  We take a dart, and throw it at the wall. The probability that our dart would land within the central half of a SNR on the wall is analogous to the probability of the central half of a SNR being on top of a BH.  This is not very dependent on the number of BHs in our sample.  However, if we increase the number of SNRs (and the area on the wall taken up by SNRs), then this probability increases. \par
Now, imagine the wall has dots on it (locations of galactic BHs). Now, we take a sticker and throw it at the wall. The probability that the central half of the sticker would land on a dot on the wall is analogous to the probability of a BH being seen in the central half of a SNR. If we increase the number of SNRs in the galaxy, this probability would not change. If we increase the number of BHs (dots on the wall), then this probability of overlap increases. \par
Based on our simulations, we can report the 3$\sigma$ upper limits on these probabilities. The probability of finding a given BH inside the central half of the SNR of $P < 0.37\%$ and a probability on finding the central half of a given SNR coincident with a BH is $P < 1.695\%$. While we cannot definitively rule out a coincidental overlap in the case of G351.9$-$0.9 and Swift J1728.9$-$3613, we have determined that the probability of overlap is low. As mentioned in Section \ref{sec:prob}, it is certain we have only observed a fraction of the population of BHs and SNRs in the MW. However, any observational biases in our simulations are reflected in the observational biases that affect the detection of these objects.
\bibliography{ref}{}
\bibliographystyle{aasjournal}




\end{document}